\documentclass[onecolumn,sort&compress,numbers]{els-mrw} 

\usepackage{amsmath,amssymb,amsfonts,amsthm,makeidx,graphicx}
\usepackage{txfonts}
\usepackage{helvet}
\newcommand{\bd}{\begin{displaymath}}
\newcommand{\ed}{\end{displaymath}}
\newcommand{\be}{\begin{equation}}
\newcommand{\ee}{\end{equation}}
\newcommand{\bear}{\begin{eqnarray}}
\newcommand{\eear}{\end{eqnarray}} 
\newcommand{\ba}{\begin{array}}
\newcommand{\ea}{\end{array}}

\newcommand{\del}{\partial}
\newcommand{\nin}{\noindent}
\newcommand{\vs}{\vskip0.2cm}
\newcommand{\cL}{{\cal L}}

\newcommand{\Tr}{\rm Tr}

\newcommand{\dslcv}{\hskip-0.1cm\not\hskip-0.13cm}

\newcommand{\grad}{\vec{\bigtriangledown}}
\newcommand{\vf}{\mathbf}

\begin{document}


\chapter{Spontaneous Symmetry Breaking and the Higgs Mechanism }\label{chap1}

\author[1]{Gustavo Burdman}%


\address[1]{\orgname{University of Sao Paulo}, \orgdiv{Department of
    Mathematical Physics}, \orgaddress{R. do Matao 1371, Sao Paulo, SP Brazil}}

\articletag{Chapter Article tagline: update of previous edition, reprint.}

\maketitle

\begin{abstract}[Abstract]
The Higgs sector of the standard model of particle physics plays a
central role in the generation of all the masses of elementary
particles known so far. Here we give a pedagogical introduction  to
all the elements leading ot the Higgs mechanism and the Higgs boson,
starting with the spontaneous symmetry breaking of global symmetries
and the Goldstone theorem. We then consider the case of gauge
symmetries, i.e. the Higgs mechanism, and its application to the
electroweak sector of the standard model.
We close with a reflection on the possible open questions that the
very introduction of the Higgs sector in the standard model posses. 
\end{abstract}

\begin{keywords}
 	symmetry breaking\sep Higgs mechanism\sep Higgs boson \sep
        Nambu-Goldstone boson \sep standard model
\end{keywords}



\section*{Objectives}
\begin{itemize}
	\item An introduction of the concept of spontaneous breaking
          of  a ccontinuous symmetry
	\item Explore the different cases for abelian and non
          abelian symmetries.
          \item Introduce Goldstone's theorem and its
            consequences in the particle spectrum. 
	\item Introduce the Higgs mechanism using various examples. 
	\item Apply all what was introduce above to the eleectroweak
          standard model.
\end{itemize}

\section{Introduction}
\label{intro}

The Higgs mechanism plays a central role in the standard model (SM) of
particle physics and, therefore, in our understanding of fundamental
physics. It is part of the very general phenomenon of spontaneous
symmetry breaking, as applied to {\em gauge} symmetries. We can trace the
beginings of these developments in a pair of papers by Yoichiro Nambu,
Refs.~\cite{Nambu:1961tp,Nambu:1961fr},
where it is shown that a
model inspired by the BCS theory of
superconductivity \cite{Bardeen:1957kj} can be used to explain the
nucleon mass and requires a zero-mass pion. This mechanism is
later identified as the spontaneous breaking of a {\em global}
symmetry by Goldstone's
theorem~\cite{Goldstone:1961eq,Goldstone:1962es}, making it clear that
it applies to a large number of physical systems from the description
of superfluidity and superconductivity in  condensed matter physics,
to relativistic quantum field theories such as QCD at low energies and
the SM. In fact, the road to the Higgs mechanism, which comes together
in Refs.~\cite{Englert:1964et,Higgs:1964pj} by Brout and Englert,
Higgs and others~\cite{Guralnik:1964eu}, is first paved by the work of
Anderson~\cite{Anderson:1963pc} on the excitations of a charged
superconducting gas. The final step to put together the electroweak
SM, is taken by Weinberg~\cite{Weinberg:1967tq}, by using the Anderson,
Brout, Englert, Higgs mechanism in order to make sense of the massive
gauge bosons in Glashow's model of electroweak
unification~\cite{Glashow:1961tr}. In what follows, we present a
pedagogical review of this process, starting with the spontaneous
breaking of abelian global symmetries in Section~\ref{sec2}, followed by the generalization
for non abelian cases and a general statement of Goldstone's
theorem~in Section~\ref{sec3}. We then introduce the Anderson-Brout-Englert-Higgs mechanism, first for
abelian gauge theories in Section~\ref{sec4}, and then for the non abelian generic
case in Section~\ref{sec5}. Finally, also in Section~\ref{sec5},  we apply all of this to the construction of the
electroweak SM. In our conclusions, we summarize and reflect on the SM
Higgs sector, and how it may point to physics beyond our current
understanding of elementary particle physics.  

\section{Spontaneous Symmetry Breaking of Global Symmetries}
\label{sec2}

The Spontaneous Breaking of a Symmetry refers to a special realization
of the symmetry in question. Although the symmetry is still respected
by the full theory, i.e. the Lagrangian density still remains
invariant under symmetry transformations, the {\em ground state} of the
system is not. An example of this situation  is given by a
ferromagnet. Consider a 3D lattice of spins separated by a fixed distance
$a$ and at a temperature $T$. The (Heisenberg)  Hamiltonian of the system includes
near neighbor interactions of the type~\cite{Chaikin_Lubensky_1995}:
\begin{equation}
-J\, \vec{s_i}\cdot \vec{s_j}~,
\label{eq:sisj}
\end{equation}
with $J$ some positive constant, and  $(i,j)$ are the positions in the
lattice. The Hamiltonian of the system, the sum over all near neighbor
interactions, is clearly invariant under 3D rotations. I.e. the system
is invariant under $O(3)$ transformations. 
The internal energy $E$  is clearly  minimized by the alignment of the
spins. However, at a finite temperature $T$ there is a competing way
to minimize the free energy $F$,
\begin{equation}
  F = E - TS~,
  \label{eq:free}
\end{equation}
given by {\em maximazing} the entropy $S$. Thus, there are two
competing mechanism to minimize $F$: at ``high enough'' $T$,
increasing $S$, i.e. increasing disorder, is preferred, so there is no
macroscopic alignment. On the other hand, at ``log enough''
temperature, the first term will be more important and to minimize $F$
alignment of spins will be favored, eventually leading to a
macroscopic magnetization, $\vec{M}$. This clearly means that there is
a critical temperature $T_c$ below which the interactions result in
growing domains of magnetization and in the end, and this macroscopic
value.   
But the macroscopic  magnetization $\vec{M}$ points in a given
arbitrary direction. This clearly is not invariant under
rotations. So we see that, although the microscopic Hamiltonian
remains $O(3)$ invariant, the ground state of finite $\vec{M}$ is not!
This is what we call spontaneous symmetry breaking of the $O(3)$
rotation symmetry in a ferromagnet. If we imagine reheating the
material above $T_c$ and then lowering the temperature below it a
large number of times, each the macroscopic magnetization will point
in a different arbitrary direction.  In the limit of infinitely many
of these experiments, the resulting magnetization vectors $\vec{M}$
will form the surface of a sphere with their tips. This is because
there are infinitely many possible ground states all with the same
energy, such that all of them combined restore the symmetry
$O(3)$. However, each time the ground state
is obtained it is not invariant. The expansion of fluctuations around
the ground state with the same energy but with $\vec{M}$ fluctuating
away from the ground state position signals the presence of {\em
  gapless} excitations, in this case spin waves of very long (actually
infinite) wavelength. As we will see below, this is a general feature
of the spontaneous breaking of continuous symmetries. Since the ground
state about which we expand the fluctuations is continuously
degenerate with all the other possible outcomes of $\vec{M}$, it cost
no energy to excite this particular fluctuations, i.e they are
gapless.

Another  example of spontaneous symmetry
breaking (SSB) is encountered in Bose-Einstein condensation and the
phenomenon of superfluidity~\cite{Girvin_Yang_2019}. In this case, the symmetry is a global
$U(1)$. We will examine this case in detail below. In particular we
will see that the gapless excitation in the relativistic formulation
appears as a massless particle. 
But before we go into this particular example, we start by stating the
generic situation of SSB, a simple quantum mechanical derivation of
what we would later call Goldstone's theorem. 

\subsection{Spontaneous Breaking and Gapless Excitations} 

\nin
Noether's theorem tells us that for each continuous symmetry in the
Lagrangian $\cL(\phi,\del_\mu\phi)$  there is a conserved current
$J^\mu$, i.e.
 \vs\be
\del_\mu J^\mu = 0~.
\label{jconserved1}
\ee\vs\nin 
We can restate this by saying that the charge associated with this symmetry
\vs\be
Q = \int d^3x \,J^0~,
\label{chargedef}
\ee\vs\nin
is conserved. This is easily checked by computing
\vs\be
\frac{d Q}{dt} = \int d^3x\,\del_0 J^0 = \int d^3x \,\grad\cdot\vf J =
\int_{S_{\infty}} d\vf s\cdot\vf  J =0~,
\label{dqdtzero} 
\ee\vs\noindent
where in the last step we assume there are no sources at infinity.

\nin
Now, in the presence of a continuous symmetry, quantum states
transform under the symmetry as
\vs\be
|\psi\rangle \to e^{i\alpha Q}\,|\psi\rangle ~,
\label{psitransf1}
\ee\vs\nin 
where $\alpha$ is a real constant, i.e. a continuous parameter. In
particular, if the ground state is invariant under the symmetry this
means that
\vs\be
|0\rangle \to e^{i\alpha Q} |0\rangle = |0\rangle~,
\label{gsinvariant}
\ee\vs\nin
with the last equality implying 
\vs\be
Q |0\rangle =0~.
\label{qkillvacuum}
\ee\vs\nin
In other words, if the ground state is invariant under a continuous
symmetry the associated charge $Q$ annihilates it. This is the normal
realization of a symmetry. 
But if 
\vs\be
Q|0\rangle \not=0~,
\label{gsnotinvariant}
\ee\vs\nin
then this means that 
\vs\be
|0\rangle \to e^{i\alpha Q} |0\rangle  \equiv |\alpha\rangle\not=|0\rangle~,
\label{alphadef1}
\ee\vs\nin
where we defined the states $|\alpha\rangle$ by the continuous 
parameter of the transformation connecting it to the ground state. 
In general, this is the situation when a symmetry is broken. But it is
possible to have (\ref{gsnotinvariant}) and still have a conserved
charge. In other words to have 
\vs\be
\frac{dQ}{dt} = 0~.
\label{dqdtstillzero}
\ee\vs\nin
Having both (\ref{gsnotinvariant}) and (\ref{dqdtstillzero}) satisfied
at the same time corresponds to what we call spontaneous symmetry
breaking (SSB): the charge is still conserved, but the ground state is not
invariant under a symmetry transformation. 
\vs\be\boxed{
\left(Q|0\rangle \not=0, \quad \frac{dQ}{dt} =0\right)\Rightarrow
{\rm SSB}}~.
\label{thisisssb}
\ee\vs\nin
This is what happens in a ferromagnet below a critical
temperature, as discussed above in the introduction to this section.

Since the charge is conserved we have that $[H,Q]=0$. Then, 
given a Hamiltonian $H$ acting on a state $|\alpha\rangle$ connected
to the ground state, we can write
\vs\bear
H|\alpha\rangle &=& H e^{i\alpha Q}|0\rangle = e^{i\alpha Q} H|0\rangle
= E_0 e^{i\alpha Q}|0\rangle\nonumber\\
&=& E_0 |\alpha\rangle~.
\label{eofalpha}
\eear\vs\nin
So we conclude that (\ref{thisisssb}) results in a continuous family
of degenerate states $|\alpha\rangle$ with the same energy of the
ground state, $E_0$. Going from the ground state $|0\rangle$ to the
$|\alpha\rangle$ states costs no energy. These are the gapless states
characteristic of SSB. They are the Nambu-Goldstone modes. In a
relativistic quantum field theory they correspond to massless
particles, as we will see in the following example.

\nin
\subsection{Spontaneous Breaking of a Global $U(1)$ Symmetry}

\nin
We will consider a complex scalar field, the simplest system to
illustrate the spontaneous breaking of a global symmetry and the
appearance  of massless particles. This is the relativistic version of
the  superfluid. The Lagrangian is 
\vs\be
\cL = \frac{1}{2} \del_\mu\phi^* \del^\mu\phi
-\frac{1}{2}\mu^2\phi^*\phi -\frac{\lambda}{4}
\left(\phi^*\phi\right)^2~.
\label{cxscalar1}
\ee\vs\nin
As we well know, $\cL$ is invariant under the $U(1)$ symmetry
transformations
\vs\be
\phi(x)\to e^{i\alpha} \phi(x)~,\qquad \phi^*(x) \to e^{-i\alpha}
\phi^*(x)~,
\label{u1forphi}
\ee\vs\nin
where $\alpha$ is a real constant. Here the $U(1)$ symmetry is
equivalent (isomorphic)  to a rotation in the complex plane defined by
\vs\be
\phi(x) = \phi_1(x) + i\phi_2(x)~,\qquad \phi^*(x) = \phi_1(x)
-i\phi_2(x)~,
\label{realimg}
\ee\vs\nin
where $\phi_{1,2}(x)$ are real scalar fields. Then we see that  $U(1) \sim O(2)$. For instance, had we started with a purely
real field $\phi(x) = \phi_1(x)$, i.e. $\phi_2(x)=0$, the $U(1)$
transformations (\ref{u1forphi}) would result in 
\vs\be
\phi(x) = \phi_1(x) \to \cos\alpha \phi_1(x) + i \sin\alpha\phi_1(x)~,
\label{rotation}
\ee\vs\nin
as illustrated in Figure~\ref{fig:fig1} below.
\begin{figure}[t]
	\centering
	\includegraphics[width=.4\textwidth]{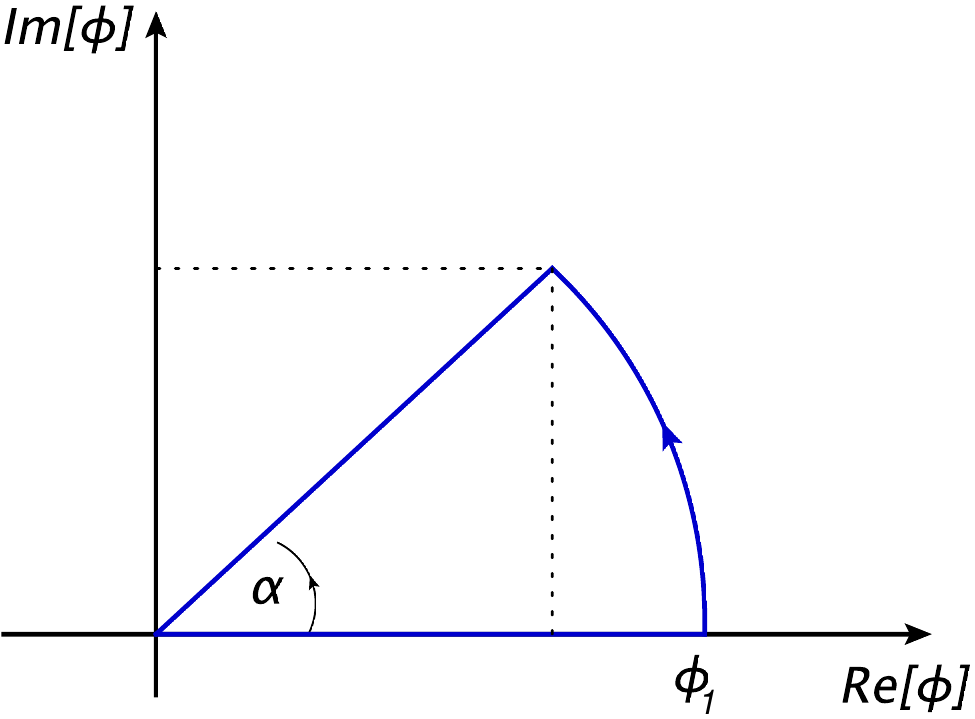}
	\caption{The $U(1)$ rotation $\phi\to e^{i\alpha}\phi$ for an initially real
field. }
	\label{fig:fig1}
\end{figure}
We now consider the (classical) potential
\vs\be
V = \frac{1}{2}\mu^2\phi^*\phi +
\frac{\lambda}{4}\left(\phi^*\phi\right)^2~.
\label{potential1}
\ee\vs\nin    
For $\mu^2>0$ $V$ has a minimum at $(\phi^*\phi)_0=0$. On the other
hand, if $\mu^2<0$ there is a non trivial minimum for $\lambda>0$
resulting from the competition of the first and second terms in
(\ref{potential1}). Redefining 
\vs\be
\mu^2 \equiv -m^2~,
\label{mutomdef}
\ee\vs\nin
with $m^2>0$, the minimum of the potential now is
\vs\be
\left(\phi^*\phi\right)_0 = \frac{m^2}{\lambda} \equiv v^2~.
\label{vevdef1}
\ee\vs\nin
Here $v^2$ is the expectation value of the $\phi^*\phi$ operator in
the ground state, i.e.
\vs\be
\langle 0| \phi^*\phi |0\rangle =v^2~.
\label{vdef2} 
\ee\vs\nin
The resulting potential, known as the mexican hat,  is shown in Figure~\ref{fig:fig2} below. 
\vs
\begin{figure}[h]
  \centering
  \includegraphics[width=.7\textwidth]{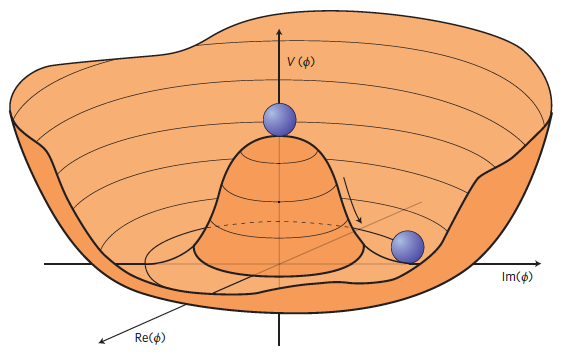}
  \caption{
The mexican hat potential for the $\mu^2<0$ case in
Eq.~(\ref{potential1}). The condition (\ref{vdef2}) corresponds to
the circle at the botton of the potential, the circle of radius $v$.
}
\label{fig:fig2}
\end{figure}
\nin    
The radius is fixed through
\vs\be
(\phi^*\phi)_0 =v^2 = \phi_1^2 +\phi_2^2~,
\label{circle1}
\ee\vs\nin
but the phase is undetermined. We need to fix it in order to choose a
ground state to expand around. Any choice should be equivalent
\vs\bear
\langle\phi_1\rangle &=&v\qquad\qquad\langle\phi_2\rangle =0\nonumber\\
\langle\phi_1\rangle &=&\frac{v}{\sqrt{2}}  \qquad\quad\langle\phi_2\rangle
  =\frac{v}{\sqrt{2}}\nonumber\\
\vdots &~&\qquad\qquad \vdots \nonumber\\
\langle\phi_1\rangle& =&0 \qquad\qquad\langle\phi_2\rangle =v\nonumber~.
\eear\vs\nin
This particular choice is what constitutes spontaneous symmetry
breaking. These points correspond to choosing {\em one point} in the
circle at the botton of the potential in Figure~\ref{fig:fig2}. 
\vs
\begin{figure}[h]
  \centering
  \includegraphics[width=.7\textwidth]{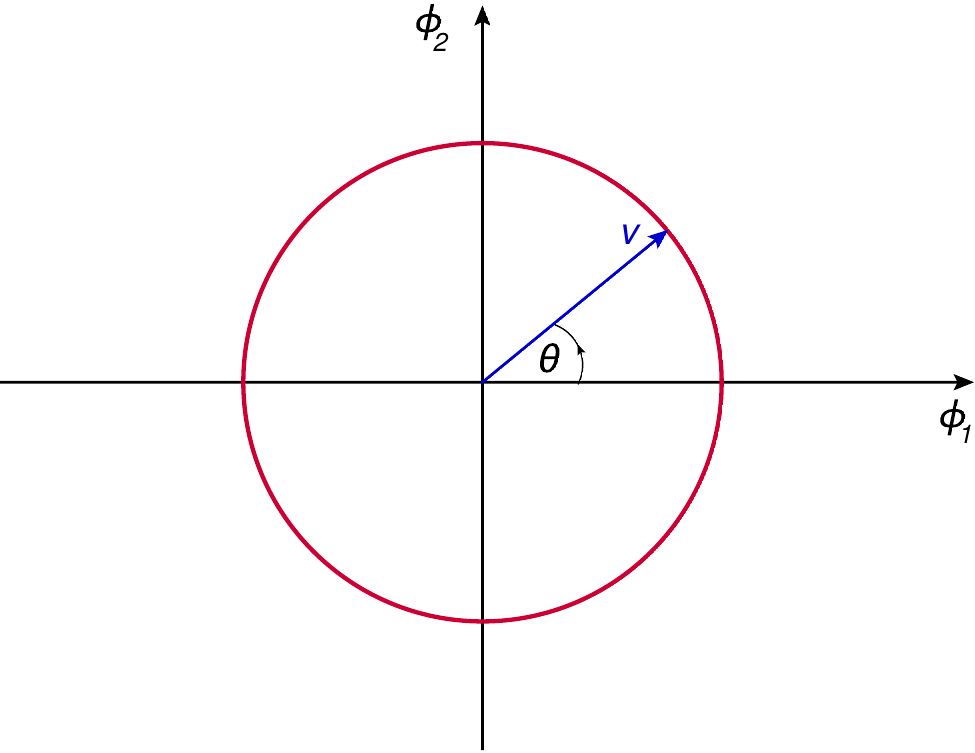}
  \caption{
The red circle represents the locus points of the minimum of the
potential (\ref{potential1}) for $\mu^2<0$. The radius is $v$, a real
number. The phase is not determined by the minimization.  Here
$phi_1=Re[\phi$ and $\phi_2=Im[\phi]$.
}
\label{fig:fig3}
\end{figure}
We need to fix the phase $\theta=\theta_0$ arbitrarily in
order to expand around {\em this}  ground state. For instance, let us
choose  $\langle\phi_1\rangle= Re[\phi]=v$, and
$\langle\phi_2\rangle=Im[\phi]=0$. This allows us to expand the field $\phi(x)$
around this ground state as
\vs\be
\phi(x) = v + \eta(x) + i\xi(x)~,
\label{expansion1}
\ee\vs\nin
where $\eta(x)$ and $\xi(x)$ are {\em real} scalar fields statisfying
\vs\be
\langle 0 | \eta(x)|0\rangle =0,\qquad\qquad \langle 0|
\xi(x)|0\rangle =0~.
\label{zerovevs}
\ee\vs\nin
This obviously corresponds to $\phi_1(x)=v+i\eta(x)$ and
$\phi_2(x)=\xi(x)$. We can now rewrite the Lagrangian (\ref{cxscalar1})
in terms of $\eta(x)$ and $\xi(x)$. This is
\vs\bear
\cL &=& \frac{1}{2}\del_\mu\eta\del^\mu\eta
  +\frac{1}{2}\del_\mu\xi\del^\mu\xi +\frac{1}{2}m^2
  \left(v+\eta-i\xi\right)\left(v+\eta+i\xi\right) \nonumber\\
&~&-\frac{\lambda}{4}\left[\left(v+\eta-i\xi\right)\left(v+\eta+i\xi\right)\right]^2~,
\label{linetaxi1}
\eear\vs\nin
where we used  (\ref{mutomdef}).  Using (\ref{vevdef1}) and focusing
on the terms quadratic in the fields, we obtain
\vs\be
\cL = \frac{1}{2}\del_\mu\eta\del^\mu\eta
  +\frac{1}{2}\del_\mu\xi\del^\mu\xi -m^2\eta^2 + {\rm
    ~~interactions}~.
\label{noximass1}
\ee\vs\nin
So we see that when we expand around the ground state defined by
(\ref{expansion1}) we end up with a theory of a real scalar field with
mass ($\eta$) and a massless state $\xi$. That is 
\vs\be
m_{\eta} = \sqrt{2} m,\qquad\qquad m_\xi =0~.
\ee\vs\nin
This result is a reflection of Goldstone's theorem: a spontaneously
broken continuous symmetry, here a $U(1)$, results in massless
states. The massless state, $\xi(x)$ is called a Nambu-Goldstone boson (NGB).

Notice that the result would be exactly the same had we chosen
any other angle in Figure~\ref{fig:fig3} instead of $\theta=0$. One
simple way to check this is to use a different parametrization of
$\phi(x)$. We write
\vs\be
\phi(x) \equiv \left[v + h(x)\right] e^{i\pi(x)} ~,
\label{radial1}
\ee\vs\nin 
where $h(x)$ and $\pi(x)$ are real scalar fields, also satisfying 
\vs\be
\langle 0|h(x)|0\rangle = 0,\qquad\qquad \langle 0|\pi(x)|0\rangle =0~.
\ee\vs\nin
Then from (\ref{radial1}) it is pretty obvious that $\pi(x)$ does not
enter in the potential, and therefore will not have a mass term. It is
very simple to obtain the Lagrangian (\ref{cxscalar1}) in terms of
$h(x)$ and $\pi(x)$ using (\ref{radial1}). This is 
\vs\be
\cL =\frac{1}{2} \del_\mu h\del^\mu h +
\frac{1}{2}\del_\mu\pi\del^\mu\pi -m^2 h^2 + {\rm ~~interactions}~,
\label{nopimass1}
\ee\vs\nin
which is exactly the same theory as the one in (\ref{noximass1}),
i.e. a massive state with $m_h=\sqrt{2}m$ and a massless particle,
here the $\pi(x)$, the NGB. Thus, we see that independently of the
parametrization chosen, we end up with the same field theory~\cite{Donoghue_Golowich_Holstein_2023}.  

\nin
We will now see a more general  derivation of Goldsone's theorem
for non Abelian symmetries. 
broken symmetry.

\section{The Case of Non Abelian Global Symmetries}
\label{sec3}

We start with the lagrangian for a scalar field $\phi$,
\vs\be
\cL = \del_\mu\phi^\dagger \del^\mu\phi -\frac{\mu^2}{2}
\phi^\dagger\phi -\frac{\lambda}{4}\Bigl(\phi^\dagger\phi\Bigr)^2~.
\label{lagscalar1}
\ee\vs\nin
The lagrangian above is invariant under the transformation
\vs\be
\phi(x) \to e^{i \alpha^a\,t^a}\,\phi(x)~,
\label{globalt1}
\ee\vs\nin
where the $t^a$ are the generators of the non abelian group $G$, and
the arbitrary parameters $\alpha^a$ are constants.
Here the scalar field $\phi(x)$ must carry a group index in order for
(\ref{globalt1}) to make sense. 
\nin
We say the  symmetry is spontaneously broken  if
we have
\vs\be
\mu^2=-m^2 < 0~.
\ee\vs\nin
Then the potential has a non trivial minimum at
\vs\be
\Bigl(\phi^\dagger\phi\Bigr)_0 =\langle\phi^\dagger \phi\rangle =
\frac{m^2}{\lambda}\equiv v^2~.
\label{minimum1}
\ee\vs\nin
However, we need to ask {\em how} is the symmetry spontaneously
broken. In other words, Spontaneous Symmetry Breaking (SSB) means that
the value of the field at the minimum, let us call it the vacuum
expectation value (VEV) of the field $\langle\phi\rangle$ , is not invariant under the
symmetry transformation (\ref{globalt1}). That is,
\vs\be
\langle\phi\rangle \to e^{i\alpha^a t^a}\,\langle\phi\rangle = \Bigl(1
+ i\alpha^a t^a +\cdots\Bigr)\,\langle\phi\rangle~,
\ee\vs\nin
can be either equal to $\langle\phi\rangle$ or not. 
This tells us that if 
\vs\be
t^a\langle\phi\rangle =0~,
\label{unbroken1}
\ee\vs\nin
the ground state is invariant under the action of the symmetry ({\em
  unbroken symmetry directions }), whereas if
\vs\be
t^a\langle\phi\rangle \not=0~,
\label{broken1}
\ee\vs\nin
the ground state is not invariant ({\em broken symmetry
  directions}). We see that some of the generators will annihilate the
ground state $\langle\phi\rangle$, such as in (\ref{unbroken1}),  whereas
others will not.
In the first case, these directions in group space will correspond to
preserved or unbroken symmetries. Therefore, there should not be
massless NGBs associated with them. On the other hand, if the
situation is such as in (\ref{broken1}), then the ground state is not
invariant under the symmetry transformations {\em defined by these
  generators}. These directions in group space defined {\em broken
  directions or generators} and there should be a massless NGB 
associated with each of them. Thus, as we will see in more detail
below, the number of NGBs will correspond to the total number of
generators of G, minus the number of unbroken generators, i.e. the
number of {\em broken generators}.

\subsection{Example 1: $SU(2)$}

\nin
As a first example, let us consider the case where the symmetry
transformations are those associated with the group $G=SU(2)$. The
{\em three} generators of $SU(2)$ are
\vs\be
t^a = \frac{\sigma^a}{2}~,
\label{gensofsu2}
\ee\vs\nin
with ${\sigma^a}$ the three Pauli matrices. This means that the scalar
fields appearing in the lagrangian (\ref{lagscalar1}) are {\em
  doublets} of $SU(2)$, i.e. we can represent them by a column vector
\vs\be
\phi(x) = \left(\ba{c}
  \phi_1(x)\\
  ~\\
  \phi_2(x)\ea\right)
\label{fundrep}~,
\ee\vs\nin
and that the symmetry transformation can be written as\footnote{We put
  the group indices in the fields upstairs for future notational
  simplicity. There is no actual meaning to them being ``up'' or
  ``down'' indices, but the summation convention still holds.}
\vs\be
\phi^i(x) =\Bigl(\delta^{ij}  + i \alpha^a
t^a_{ij}+\cdots\Bigr)\,\phi^j(x)~,
\label{globalt2}
\ee\vs\nin
where $i,j=1,2$ are the group indices for the scalar field in the
fundamental representation. We now need to {\em choose}  the vacuum
$\langle\phi\rangle$. This is typically informed by either the
physical system we want to describe or by the result we want to get. 
Let us choose
\vs\be
\langle \phi\rangle = \left(\ba{c} 0\\
  v\ea\right)~.
\label{vev1}
\ee\vs\nin
Clearly this satisfies (\ref{minimum1}). This choice corresponds to
having
\vs\bear
\langle {\rm Re}[\phi_1]\rangle &=&0\qquad\qquad\langle {\rm Im}[\phi_1]\rangle=0\nonumber\\
\langle {\rm  Re}[\phi_2]\rangle &=&v\qquad\qquad \langle
{\rm Im}[\phi_2]\rangle=0~,
\label{vevsofphis1}
\eear\vs\nin
in (\ref{fundrep}). We can now test what generators annihilate the
vacuum (\ref{vev1}) and which ones do not. We have
\vs\be
t^1\,\langle\phi\rangle = \frac{1}{2}\left(\ba{cc}
  0&1\\
  1&0\ea\right)\,\left(\ba{c}0\\
  v\ea\right) = \frac{1}{2}\,\left(\ba{c} v\\
  0\ea\right) \not = \left(\ba{c}0\\
0\ea\right) ~.
\ee\vs\nin
Similarly, we have
\vs\be
t^2\,\langle\phi\rangle = \frac{1}{2}\left(\ba{cc}
  0&-i\\
  i&0\ea\right)\,\left(\ba{c}0\\
  v\ea\right) = \frac{1}{2}\,\left(\ba{c} -i v\\
  0\ea\right) \not = \left(\ba{c}0\\
0\ea\right) ~,
\ee\vs\nin
and
\vs\be
t^3\,\langle\phi\rangle = \frac{1}{2}\left(\ba{cc}
  1&0\\
  0&-1\ea\right)\,\left(\ba{c}0\\
  v\ea\right) = \frac{1}{2}\,\left(\ba{c} 0\\
  -v\ea\right) \not = \left(\ba{c}0\\
0\ea\right) ~.
\ee\vs\nin
So we conclude that with the choice of vacuum (\ref{vev1}), all $SU2)$
generators are broken. This means that all the continuous symmetry
transformations generated by (\ref{globalt1}) change the chosen vacuum
$\langle \phi\rangle$. Thus, Goldstone's theorem predicts there must
be {\em three} massless NGBs. In order to explicitly see who are
these NGBs, we write the lagrangian (\ref{lagscalar1}) in terms of the real
scalar degrees of freedom as in 
\vs\be
\phi(x) = \left(\ba{c} {\rm Re}[\phi_1(x)] + i \,{\rm Im}[\phi_1(x)]\\
  ~\\
  v+ {\rm Re}[\phi_2(x)] + i\, {\rm Im}[\phi_2(x)]\ea\right)~,
\label{phiexp}
\ee\vs\nin
which amounts to expanding
about the vacuum (\ref{vev1}) as long as (\ref{vevsofphis1}) is
satisfied. Substituting in (\ref{lagscalar1}) we will find that there
are three massless states, namely, ${\rm Re}[\phi_1(x)]$ , ${\rm
  Im}[\phi_1(x)]$ and ${\rm Im}[\phi_2(x)]$, and that there is a
massive state corresponding to ${\rm Re}[\phi_2(x)]$ with a mass given
by $m$. This looks very similar to what we obtain in the
abelian case, of course. Also analogously to the abelian case, we could
have parametrized $\phi(x)$ as in
\vs\be
\phi(x) = e^{i\pi^a(x) t^a/f}\,\left(\ba{c}0\\
  v+c\,\sigma(x)\ea\right)~,
\label{exppar1}
\ee\vs\nin
where $\sigma(x) $ and $\pi^a(x)$, with $a=1,2,3$ are real scalar
fields, and the scale $f$ and the constant $c$ are to be determined so
as to obtain canonically normalized kinetic terms for them in
$\cL$. In fact, choosing
\vs\be
f=\frac{v}{\sqrt{2}}, \qquad\qquad c=\frac{1}{\sqrt{2}}~,
\ee\vs\nin
we arrive at
\vs\be
\cL = \frac{1}{2}\del^\mu\sigma\del_\mu\sigma
+\frac{1}{2}\del^\mu\pi^a\del_\mu\pi^a  -\frac{m^2}{2}
\Bigl(v+\frac{\sigma(x)}{\sqrt{2}}\Bigr)^2
+\frac{\lambda}{4}\, \Bigl(v+\frac{\sigma(x)}{\sqrt{2}}\Bigr)^4~,
\label{lagsigpi1}
\ee\vs\nin
from which we see that the three  $\pi^a(x)$ fields are massless and
are therefore the NGBs. Furthermore, using $m^2=\lambda v^2$, we can
extract
\vs\be
m_\sigma=m=\lambda\,v
\ee\vs\nin
The choice of vacuum $\langle\phi\rangle$ resulting in this spectrum
could have been different. For instance, we could have chosen
\vs\be
\langle\phi\rangle = \left(\ba{c}v\\0\ea\right)~.
\ee\vs\nin
But it is easy to see that this choice is equivalent to (\ref{vev1}),
and that it would result in an identical real scalar
spectrum. Similarly,
the aparently different vacuum
\vs\be
\langle\phi\rangle = \frac{1}{\sqrt{2}}\left(\ba{c}v\\v\ea\right)~,
\ee\vs\nin
results in the same spectrum.
All these vacuum choices spontaneously break $SU(2)$
{\em completely}, i.e. there are not symmetry transformations that
respect these vacua. Below we will see an example of partial
spontaneous symmetry breaking.

\subsection{Example 2: $SU(3)$}

If we now consider that $\cL$ in (\ref{lagscalar1})
is invariant under $SU(3)$ global transformations, there are going to
be  $3^2-1=8$ generators. A convenient basis for them is provided by
the Gellmann matrices:
\vs\bear
t^1&=& \frac{1}{2}\left(\ba{ccc}
  0&1&0\\
  1&0&0\\
  0&0&0\ea\right)
\qquad\qquad
t^2= \frac{1}{2}\left(\ba{ccc}
  0&-i&0\\
  i&0&0\\
  0&0&0\ea\right)\nonumber\\
t^3&=& \frac{1}{2}\left(\ba{ccc}
  1&0&0\\
  0&-1&0\\
  0&0&0\ea\right)
\qquad\qquad
t^4= \frac{1}{2}\left(\ba{ccc}
  0&0&1\\
  0&0&0\\
  1&0&0\ea\right)\nonumber\\
t^5&= &\frac{1}{2}\left(\ba{ccc}
  0&0&-i\\
  0&0&0\\
  i&0&0\ea\right)
\qquad\qquad
t^6= \frac{1}{2}\left(\ba{ccc}
  0&0&0\\
  0&0&1\\
  0&1&0\ea\right)\nonumber\\
t^7&=& \frac{1}{2}\left(\ba{ccc}
  0&0&0\\
  0&0&-i\\
  0&i&0\ea\right)
\qquad\qquad
t^8= \frac{1}{\sqrt{3}}\left(\ba{ccc}
  1&0&0\\
  0&1&0\\
  0&0&-2\ea\right)~,\label{su3gens1}
\eear\vs\nin
which are nothing but ``stretched'' Pauli matrices. In fact, it is
useful to note that for $i=1,2,3$ we have
\vs\be
t^i=\frac{1}{2}\left(\ba{cc}
  \sigma^i&0\\
  0\,\,\,0&0\ea\right)~,
\label{su3pauli}
\ee\vs\nin
which means that $SU(2)$ is a subgroup of $SU(3)$, since its
generators appears as block diagonal matrices in the generators of
$SU(3)$ .

\nin
Let us now consider the following choice of vacuum
\vs\be
\langle\phi\rangle = \left(\ba{c}
  0\\
  0\\
  v\ea\right)~,
\ee\vs\nin
which clearly satisfies the constraint $\langle\phi^\dagger\phi\rangle
=v^2$.
It is straightforward to verify that the generators in
(\ref{su3pauli}) annihilate this vacuum, i.e.
\vs\be
t^i\,\langle\phi\rangle = \left(\ba{c}0\\0\\0\ea\right)~,
\ee\vs\nin
for $i=1,2,3$. This means that with this vacuum choice,
the $SU(2)$ subgroup of $SU(3)$ leaves the vacuum invariant, i.e. this
subgroup {\em is not} spontaneously broken.
On the other hand, we can see that
\vs\be
t^{4,\cdots,8}\,\langle\phi\rangle \not=\left(\ba{c}0\\0\\0\ea\right)~,
\ee\vs\nin
resulting in $5$ broken generators. Then there are $5$ continuous
symmetries spontaneously broken, resulting in $5$ NGBs. This is what
we call partial spontaneous breaking: the breaking pattern is
\vs\be
SU(3)\to SU(2)~,
\ee\vs\nin
from which we can see that the number of NGBs can also be thought of
as the number of original generators, minus the number of generators
of the unbroken group.

\subsection{Goldstone Theorem }

\nin
We go back to considering the infinitesimal transformation
(\ref{globalt2}), but we rewrite it as
\vs\be
\phi^i \to \phi^i + \Delta^i(\phi)~,
\label{globalt3}
\ee\vs\nin
where we defined
\vs\be
\Delta^i(\phi) \equiv i \alpha^a\,(t^a)_{ij}\phi^j~.
\label{deltadef1}
\ee\vs\nin
If the potential has a non trivial minimum  at $\phi^i(x)=\phi^i_0$, then it is satisfied that
\vs\be
\frac{\del V(\phi^i)}{\del\phi^i}\Bigr|_{\phi_0} =0~.
\label{der1iszero}
\ee\vs\nin
We can then expand the potential around the minimum as
\vs\be
V(\phi^i) = V(\phi^i_0)
+\frac{1}{2}\,\Bigl(\phi^i-\phi^i_0\bigr)\,\Bigl(\phi^j-\phi^j_0\Bigr)\,\frac{\del^2
  V}{\del\phi^i\del\phi^j}\Bigr|_{\phi_0}
+\cdots~,
\label{taylorexp1}
\ee\vs\nin
where the first derivative term is omitted in light of
(\ref{der1iszero}). The second derivative term in (\ref{taylorexp1})
above defines a matrix with units of square masses:
\vs\be
M^2_{ij}\equiv \frac{\del^2
  V}{\del\phi^i\del\phi^j}\Bigr|_{\phi_0} \geq 0~.
\label{msqdef1}
\ee\vs\nin
where the last inequality results from the fact that  $\phi^0$ is a
minimum. $M^2_{ij}$ is the mass squared matrix.
We are now in the position to state Goldstone's theorem in this
context.
\vs\vs\vs\vs\nin
\underline{\bf Theorem}: \\
``For each symmetry of the lagrangian that {\em is
  not } a symmetry of the vacuum $\phi_0$, there is a zero eigenvalue
of $M^2_{ij}$ .''

\vs\vs\vs\nin
\underline{\bf Proof}: \\
The infinitesimal symmetry transformation in (\ref{globalt3}) leaves the lagrangian
invariant. In particular, it also leaves the potential invariant,
i.e.
\vs\be
V(\phi^i) = V\Bigl(\phi^i +\Delta^i(\phi)\Bigr)~.
\label{visinvariant}
\ee\vs\nin
Expanding the right hand side of (\ref{visinvariant}) and keeping
only terms leading in $\Delta^i(\phi)$, we can write
\vs\be
V(\phi^i) = V(\phi^i)  +
\Delta^i(\phi)\,\frac{\del V(\phi^i)}{\del\phi^i} ~,
  \label{expandv1}
\ee\vs\nin
which, to be satisfied  requires that
\vs\be
\Delta^i(\phi)\,\frac{\del V(\phi)}{\del\phi^i}=0~.
\label{vanish1}
\ee\vs\nin
To make this result useful, we take a derivative on both sides and
specified for $\phi^i=\phi^i_0$, i.e. we evaluate all the expression
at the minimum of the potential. We obtain
\vs\be
\frac{\del\Delta^i(\phi)}{\del\phi^j}\Bigr|_{\phi_0}\,\frac{\del
  V(\phi)}{\del\phi^i}\Bigr|_{\phi_0} +
    \Delta^i(\phi_0)\,\frac{\del^2
      V(\phi)}{\del\phi^j\del\phi^i}\Bigr|_{\phi_0} =0~.
 \ee\vs\nin

      But by virtue of (\ref{der1iszero}), the first term above
      vanishes, leaving us with 
      \vs\be\boxed{
\Delta^i(\phi_0)\,\,\,\frac{\del^2
  V(\phi)}{\del\phi^j\del\phi^i}\Bigr|_{\phi_0} =0}~.
\label{thisiszero}
      \ee\vs\nin
      There are two ways to satisfy (\ref{thisiszero}):

      \begin{enumerate}
      \item $\Delta^i(\phi_0) = 0$. \\
        But this means that, under a symmetry
  transformation, the vacuum is invariant, since according to
  (\ref{globalt3}) this results in
  \vs\be
\phi^i_0\to\phi^i_0~.
\ee\vs\nin

\item $\Delta^i(\phi_o)\not=0$. \\
  This requires that the second derivative factor in (\ref{thisiszero})
  must vanish, i.e.
  \vs\be
M^2_{ij}=0~.
\ee\vs\nin
We then conclude that for each symmetry transformation that {\em does not
  leave the vacuum invariant} there must be a zero eigenvalue of the
mass squared matrix $M^2_{ij}$. QED.

        \end{enumerate}

\section{The Higgs Mechanism: Spontaeous Breaking of a Gauge
  Symmetry} 
\label{sec4}%

We have seen that the spontaneous breaking of a continuous symmetry
results in the presence of massless states in the spectrum, the
Nambu--Goldstone Bosons (NGB). This was the case for a
$U(1)$ global symmetry where the potential was such that the ground
state was not $U(1)$ invariant. In that case, the NGB corresponded to
the degeneracy of the ground state, i.e. it was the fluctuation going
around the degenerate minimum and as such it corresponded to a
massless state. 

We 
will now consider the situation when the $U(1)$ symmetry studied earlier
is gauged. That is, is a local $U(1)$ symmetry such as for example in
QED. As we will soon see, the consequences for the spectrum when the
spontaneously broken symmetry is gauged are drastic.
We start with the lagrangian of a scalar field charged under a gauged
$U(1)$ symmetry just as QED. This is given by
\vs\be
\cL = \frac{1}{2} (D_\mu\phi)^* D^\mu\phi -V(\phi^*\phi) -\frac{1}{4}
F_{\mu\nu}F^{\mu\nu}~,
\label{lag1} 
\ee\vs\nin 
where the covariant derivative is defined by
\vs\be
D_\mu \phi = (\del_\mu +ieA_\mu)\phi~,
\label{covderdef}
\ee\vs\nin
and the scalar and gauge field transformations under the $U(1)$ gauge
symmetry are 
\vs\bear
\phi(x) &\to & e^{i\alpha(x)} \phi(x) \nonumber\\
&~&\label{fieldtransf1}\\
A_\mu(x) &\to & A_\mu(x) -\frac{1}{e}\del_\mu\alpha(x)~.\nonumber
\eear\vs\nin
Finally, the gauge field $A_\mu(x)$ has a kinetic term given by the
square of the  gauge invariant field strength as usual
\vs\be
F_{\mu\nu} = \del_\mu A_\nu - \del_\nu A_\mu~.
\label{fmunudef}
\ee\vs\nin 
With (\ref{covderdef}), (\ref{fieldtransf1}) and (\ref{fmunudef}) the
lagrangian in (\ref{lag1}) is clearly gauge invariant. 

\nin
In order to implement spontaneous breaking we choose the potential as
\vs\be
V(\phi^*\phi) = \frac{1}{2}\mu^2\phi^*\phi +
\frac{\lambda}{4}\left(\phi^*\phi\right)^2~,
\label{pot1}
\ee\vs\nin
which is the same form we used for the breaking fo the global $U(1)$
and corresponds to the only renormalizable terms allowed by the
symmetry in four spacetime dimensions. What follows next pertaining
the minimum of the potential is identical to what we saw for the
global symmetry case. If $\mu^2>0$ the minimum of $V$ in (\ref{pot1})
is $\phi=0$. However if $\mu^2<0$ then we rewrite the potential as
\vs\be
V(\phi^*\phi) = -\frac{1}{2}m^2\phi^*\phi +
\frac{\lambda}{4}\left(\phi^*\phi\right)^2~,
\label{pot2}
\ee\vs\nin
where we have defined the positive constant $m^2=-\mu^2$.
Repeating the expansion in (\ref{expansion1}) in Section~\ref{sec3}
we obtain once again
\vs\bear
m_\eta &=& \sqrt{2} m = \sqrt{2\lambda} v\nonumber\\
& ~&\label{spectrum}\\
m_\xi&=& 0~.\nonumber
\eear\vs\nin
And once again, we identify $\xi(x)$ with the massless NGB. The difference with
respect to the SSB of a global $U(1)$ comes in when we look at what
happens in the scalar kinetic term. 
This is 
\vs\bear
\frac{1}{2} (D_\mu\phi)^*D^\mu\phi &=&
\frac{1}{2}\del_\mu\eta\del^\mu\eta +
\frac{1}{2}\del_\mu\xi\del^\mu\xi +\frac{1}{2} \,e^2\,v^2\,A_\mu A^\mu
\nonumber\\
& ~& \label{kinscal1}\\
&&+ \,e\, v\, A_\mu\,\del^\mu\xi + \cdots~,
\nonumber
\eear\vs\nin
where we have explicitly written the terms quadratic in the fields,
and the dots denote interactions terms that are cubic or quadratic in
them. Besides the kinetic terms for $\eta(x)$ and $\xi(x)$ we notice
two terms. The first one is an apparent gauge boson mass term. It
implies that the gauge boson has acquired a mass given by
\vs\be
m_A = e\,v~.
\label{amass1}
\ee\vs\nin
However, this does not mean that the gauge symmetry is not been
respected. In fact, all we have done with respect to (\ref{lag1} )
is to expand the theory around the ground state in terms of fields
that have zero expectation values there. In other words, we just
performed a change of variables. However, the fact the we are
expanding the theory around a minimum that {\em does not}  respect the
symmetry is resulting in a mass for the gauge boson. This means that
the gauge symmetry has been {\em spontaneously} broken. But since we
have not added any terms that violated explicitly the $U(1)$ gauge
symmetry, the symmetry {\em has not} been {\em explicitly} broken and
therefore currents and charges must still be conserved. We will go
into this point in more detail later. 

\begin{figure}
  \begin{center}
    \includegraphics[width=.4\textwidth]{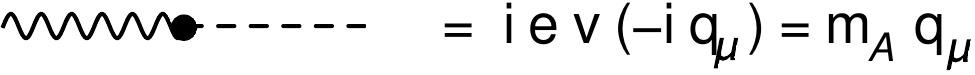}
\caption{
Feynman rule for the non-diagonal contribution to the two-point
function in (\ref{kinscal1}).
}
\label{fig:fig4}
\end{center}
\end{figure}

\nin
The second notable aspect in (\ref{kinscal1}) is the term mixing the
gauge boson with the $\xi(x)$ field, the would-be NGB. Having a term
like this, i.e. a non-diagonal contribution to the two-point function, implies that we have
to include a Feynman diagram as the one in
Figure~\ref{fig:fig4}. Although in principle there is no problem with
having a non-diagonal Feynman rule such as this as long as we always
remember to include it, it is interesting to see how to diagonalize it
and what are the consequences of doing that. The idea is to choose a
gauge for $A_\mu(x)$ such that we can cancel this term once we go to
the new gauge. The theory has to be physically equivalent to the one
with (\ref{kinscal1}). Choosing a specific gauge corresponds to
choosing a scalar function $\alpha(x)$ in the gauge tranformations
(\ref{fieldtransf1}). In particular, if we choose
\vs
\nin    

\vs\be
\alpha(x) = -\frac{1}{v}\,\xi(x)~,
\label{alphachoice}
\ee\vs\nin
we then have the gauge transformation
\vs\be
A_\mu(x) \to A'_\mu(x) = A_\mu(x) +\frac{1}{e v} \xi(x)~. 
\label{achoice}
\ee\vs\nin
Replacing $A_\mu(x)$ in terms of $A'_\mu(x)$ and $\xi(x)$ in
(\ref{kinscal1}) we have
\vs\bear
\frac{1}{2}(D_\mu\phi)^*D^\mu\phi &=& \frac{1}{2}\del_\mu\eta\del^\mu\eta +
\frac{1}{2}\del_\mu\xi\del^\mu\xi +
\frac{1}{2}e^2\,v^2\,\left(A'_\mu-\frac{1}{e v}\del_\mu\xi\right) \left(A'^\mu-\frac{1}{e v}\del^\mu\xi\right)
\nonumber\\
&~&\label{kinscal2}\\
&&+ \, e\,v\,\left(A'_\mu-\frac{1}{e v}\del_\mu\xi\right) \del^\mu\xi
+ \cdots~,\nonumber, 
\eear\vs\nin
Carefully collecting all the terms in (\ref{kinscal2}) we arrive at
the surprisingly simple expression for the scalar kinetic term:
\vs\be
\frac{1}{2}(D_\mu\phi)^*D^\mu\phi =
\frac{1}{2}\del_\mu\eta\del^\mu\eta + \frac{1}{2}\,e^2\,v^2\,A'_\mu
A'^\mu + \cdots~.
\label{kinscal3}
\ee\vs\nin
We see that the gauge boson mass term is still the same as
before. However, the $\xi(x)$ field, the massless field that we thought
would be the NGB is now gone. Its kinetic term is gone and, as we will
see later, no term with $\xi(x)$ remains in the lagrangian after this
gauge transformation. So the would-be NGB is not! When a degree of
freedom disappears from the theory just by performing a gauge
transformation, we say that this is not a physical degree of
freedom. This particular gauge without the NGB $\xi(x)$ is called the
{\em unitary gauge}, since it exposes the actual degrees of freedom of
the theory: a real scalar field $\eta(x)$ with mass $m_{\eta} =
\sqrt{2} m$ and a gauge boson with mass $m_A =e v$. In fact if we count
degrees of freedom before and after we expanded around the non-trivial
ground state, we see that before we had {\em two real scalar fields},
and {\em two degrees of freedom } corresponding to the two helicities
of a massless gauge boson, for a total of {\em four degrees of
  freedom}. But after we expanded around the ground state, we have
{\em one real scalar field}, plus {\em three polarizations} for the
now massive gauge boson, again a total of {\em four degrees of freedom}.   
It is in this sense that sometimes we say that when a gauge symmetry
is spontaneously broken, the NGB is {\em ``eaten''} by the gauge boson
to become its longitudinal polarization. This statement can be made
more precise through the {\em equivalence theorem}~\cite{Cornwall:1974km,Lee:1977eg,Chanowitz:1985hj,Veltman:1989ud}, which says that in
processes at energies much larger than $v$ (so that it does not matter
that the expectation value of the field is not zero in the ground
state) computing any observable by using the theory with a massive
gauge boson should yield the same result as using the theory with a
massless gauge boson and a massless NGB, up to corrections that go
like $v^2/E^2$, where $E$ is the characteristic energy scale of the
process in question. We will come back to  the equivalence theorem
later on when we consider the spontaneous breaking of non-abelian
gauge symmetries.

\nin
There is another, perhaps more direct, way to see that the NGB can be
{\em gauged away}, i.e. it disappears from the theory by performing a
gauge transformation. For this purpose, it is advantageous to
parametrize the scalar field not in terms of  real and imaginary
parts, but of modulus and phase. We write, as in (\ref{radial1}), 
\vs\be
\phi(x) = e^{i\pi(x)/f}\,\left(v+\sigma(x)\right)~,
\nonumber
\ee\vs\nin
where we see that this automatically satisfies that $\langle 0|\phi^*
\phi| 0\rangle =v^2$.

Using $f=v$ to get canonically normalized kinetic terms for $\pi(x)$,
we obtain 
\vs\be
m_\sigma = \sqrt{2\lambda} \,v~,
\label{sigmamass}
\ee\vs\nin
just as before.
This also means that $\pi(x)$ cannot get a mass, i.e.
\vs\be
m_\pi = 0 ~,
\label{pimass}
\ee\vs\nin
 and therefore is the
NGB. In fact, it will only appear in the lagrangian in derivative form
since it is the only way it will come down from the exponentials
before these annihilate in the kinetic scalar term.  

\nin
From the exponential parametrization (\ref{radial1}) it is also obvious how to
remove $\pi(x)$ by means of a gauge transformation. Clearly, choosing
the gauge transformation
\vs\be
\phi(x) \to \phi'(x)=e^{-i\pi(x)/v}\,\phi(x)~,
\label{gtnongb}
\ee\vs\nin
results in 
\vs\be
\phi'(x) = \left[v+\sigma(x)\right]~.
\label{nongbinphi}
\ee\vs\nin
Of course, the gauge transformation (\ref{gtnongb}) is the same we
introduced earlier in (\ref{alphachoice}) only substituting $\pi(x)$
for $\xi(x)$, and it therefore results in the same transformation for
the gauge fields as in (\ref{achoice}). Therefore, our conclusions are
exactly the same as the ones we derived by using (\ref{expansion1}) as
the field expansion: there is a massive gauge boson field with mass
$m_A=e\,v$ and a massive reals scalar with mass given by
(\ref{sigmamass}). 

\begin{figure}
  \begin{center}
    \includegraphics[width=.7\textwidth]{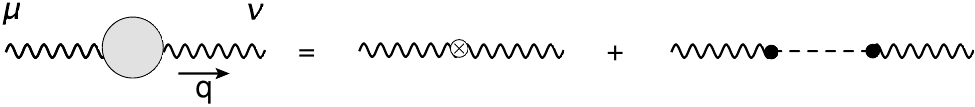}
\end{center}
\caption{
New contributions to the gauge boson two-point function at tree level
in the presence of spontaneous symmetry breaking.
The first diagram is the gauge boson mass term insertion. The second one
corresponds to the massless NGB contribution.}
\label{fig:fig5}
\end{figure}

\nin
We finally comment on the meaning of spontaneously breaking a gauge
symmetry. Specifically, we want to address the point that although the
gauge boson has acquired a mass, the gauge symmetry is still present. 
To show this, let us go back to the gauge where we have both the gauge
boson and the NGB. We want to compute the gauge boson two-point
function  at tree level. In particular we want to consider the effect
of spontaneous symmetry breaking. We will need to use the Feynman rule
illustrated in Figure~\ref{fig:fig4}. The calculation is illustrated in
Figure~\ref{fig:fig5}. 
In addition to the tree-level gauge boson propagator, there are two
new terms contributing: the gauge boson mass insertion and
the massless NGB pole. They are
\vs\bear
i\delta\Pi_{\mu\nu} &=& i m_A^2 g_{\mu\nu} + m_A q_\mu\,\frac{i}{q^2} \,m_A
(-q_\nu)\nonumber\\
&~&\label{pimunu}\\
&=& i m_A^2\left(g_{\mu\nu} -\frac{q_\mu q_\nu}{q^2}\right)~.\nonumber
\eear\vs\nin
\vs
\nin    
In the first line in (\ref{pimunu})  we used the gauge boson--NGB
mixing Feynman rule of Figure~\ref{fig:fig4}. The result is that the new
additions to the 
two-point function result to be  actually transverse. That is, we have
that 
\vs\be
q^\mu \delta\Pi_{\mu\nu} = 0~,
\label{stilltransverse}
\ee\vs\nin
so that the two-point function remains transverse, therefore
respecting the Ward identities. Since the Ward identities are
equivalent to current conservation, we conclude that the gauge
symmetry is still preserved, even in the presence of the gauge boson
mass term. We can see that this required the presence of the NGB
pole. Just having the gauge boson mass term would have resulted in a
non-transverse contribution to the two-point function, and an explicit
violation of the gauge symmetry. So having a gauge boson mass is
compatible with gauge invariance as long as it is the result of
spontaneous symmetry breaking~\cite{Weinberg_1996}.

\section{Spontaneous Breaking of a Non Abelian Gauge Symmetry}
\label{sec5}

We will now consider the case when the spontaneously broken non
abelian symmetry is gauged. As we saw for the case of abelian gauge
symmetry, the spontaneous breaking of the symmetry will be realized in
the sense of the Anderson-Higgs mechanism, i.e. the NGBs would not be
in the physical spectrum, and the gauge bosons associated with the
{\em broken } generators will acquire mass.
We will derive these results carefully in what follows\footnote{For a
  more extensive discussion see for example Refs~\cite{Peskin:1995ev}
  and \cite{Weinberg_1996}}. 

\nin
\subsection{The Generalized Anderson-Higgs Mechanism} 

\begin{figure}
  \begin{center}
    \includegraphics[width=.7\textwidth]{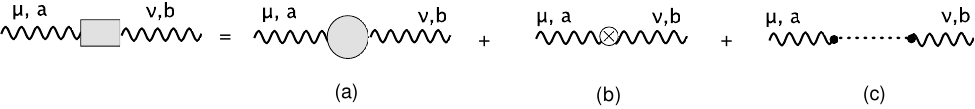}
\end{center}
\caption{Contributions to the gauge boson two point function in the
  presence of spontaneous gauge symmetry breaking. Diagram $(a)$
  includes the tree level as well as loop diagrams, all of which are
  transverse contributions. Diagram $(b)$ is the contribution from the
  gauge boson mass term. Diagram $(c)$ depicts the contribution from
  the massless NGBs.
}
\label{fig:201}
\end{figure}

\nin
We consider a lagrangian invariant under the gauge transformations
\vs\be
\phi(x)\to e^{i\alpha^a(x)\,t^a}\,\phi(x)~,
\label{gauget1}
\ee\vs\nin
where $t^a$ are the generators of the group $G$, and the gauge fields
transform as they should. 
If we consider infinitesimal gauge transfomations and write out the
field $\phi(x)$ in its groups components, we have 
\vs\be
\phi_i(x) \to \bigl(\delta_{ij} + i\alpha^a(x)
\,(t^a)_{ij}\bigr)\,\phi_j(x)
\label{gauget2}
\ee\vs\nin
In general, we consider representations where the $\phi_i(x)$ fields
in (\ref{gauget2}) are  complex. But for the purpose of our next
derivation, it would be advantegeous to consider their real
components. So if the original representation had dimension $n$, we
now have $2n$ componentes in the real fields $\phi_i(x)$. If this is
the case, then the generators in (\ref{gauget2}) {\em must be
  imaginary}, since the $\alpha^a(x)$ are real paramenter
functions. This means we can write them as
\vs\be
t^a_{ij} = i \,T^a_{ij}~,
\label{realts}
\ee\vs\nin
where the $T^a_{ij}$ are real. Also, since the $t^a$ are hermitian, we
have
\vs\be
\bigl(t^a_{ij}\bigr)^\dagger = t^a_{ij}~,
\ee\vs\nin
we see that
\vs\be
T^a_{ij} = -T^a_{ji}~,
\ee\vs\nin
so the $T^a$ are antisymmetric. In general, the lagrangian of the
gauge invariant 
theory for a scalar field in terms of the real scalar degrees of
freedom would be\footnote{Here we concentrate on the
  scalar sector of $\cL$ since it is here that SSB of the gauge
  symmetry arises. We can imagine adding fermion terms to $\cL$
  coupling them both to the gauge bosons through the covariant
  derivative, as well as Yukawa couplings between the fermions and the
  scalars. Of course, all these terms must also respect gauge invariance.} 
\vs\be
\cL = \frac{1}{2} \Bigl(D_\mu\phi_i\Bigr)\Bigl(D^\mu\phi_i\Bigr)
-V\bigl(\phi_i\bigr)~,
\label{lagrealfis}
\ee\vs\nin
where the repeated $i$ indices are summed. We can write the covariant
derivatives above as
\vs\be
D_\mu\phi(x) = \bigl(\del_\mu -ig A_\mu^a(x) t^a\bigr)\,\phi(x) = \bigl(\del_\mu + g
A^a_\mu(x) T^a\bigr)\,\phi(x)~,
\label{realcovder}
\ee\vs\nin
where we omitted the group indices for the fields and the
generators. We are interested in the situation when the potential in
(\ref{lagrealfis}) induces spontaneous symmetry breaking. To see how
this affects the gauge boson spectrum we must examine in detail the
scalar kinetic term:
\vs\bear
\frac{1}{2}\bigl(D_\mu\phi_i\bigr)\,\bigl(D^\mu\phi_i\bigr)
&=&\frac{1}{2}\del_\mu\phi_i\,\del^\mu\phi_i + \frac{1}{2} g^2 A^a_\mu
A^{b\mu} \bigl(T^a\phi\bigr)_i \bigl(T^b\phi\bigr)_i\nonumber\\
~\label{kinterm1}\\
&+& g A^a_\mu \bigl(T^a\phi\bigr)_i \,\del^\mu\phi_i~,
\nonumber
\eear\vs\nin
where we used the notation
\vs\be
\bigl(T^a\phi\bigr)_i = T^a_{ij} \phi_j~, 
\ee\vs\nin
and as usual repeated group indices $i,j$ are summed.
If the potential $V(\phi_i)$ has a non trivial minimum then,
the vacuum expectation value (VEV) of the fields $\phi_i$ at the
minimum is
\vs\be
\langle 0 |\phi_i|0\rangle =\langle\phi_i\rangle \equiv
\bigl(\phi_0\bigr)_i~,
\label{vevedef2}
\ee\vs\nin
which says that we are signling out directions in field space which
may have non trivial VEVs. Then the terms in $\cL$ quadratic in the gauge boson
fields, i.e. the gauge boson mass terms, can be readily read off
(\ref{kinterm1}): 
\vs\be
\cL_{m} = \frac{1}{2}  M^2_{ab}\,A^a_\mu A^{b\mu}~,
\label{gbmass1}
\ee\vs\nin
where the gauge boson mass matrix is defined by
\vs\be
M^2_{ab} \equiv
g^2\,\bigl(T^a\phi_0\bigr)_i\,\bigl(T^b\phi_0\bigr)_i~.
\label{massmatrixdef}
\ee\vs\nin
Since the $T^a$'s are real, the non zero eigenvalues of $M^2_{abv}$
are definite positive.  We can clearly see now that if
\vs\be
T^a\phi_0 = 0,
\label{stillgood}
\ee\vs\nin
then the associated gauge boson $A^a_\mu$ remains massless. That is,
the{\em  unbroken}  generators, which as we saw in the previous lecture, {\em
  do not have NGBs associated with them}, do not result in a mass term
for the corresponding gauge boson. On the other hand, if
\vs\be
T^a\phi_0\not=0~,
\label{notgood}
\ee\vs\nin
then we see that this results in a gauge boson mass term. The
generators satisfying (\ref{notgood}) are of course the {\em broken
  generators} which result in massless NGBs. However, just as we saw
for the abelian case, these NGBs can be removed from the spectrum by a
gauge transformation. To see how this works we consider the last term
in (\ref{kinterm1}), the mixing term. This is
\vs\be
\cL_{\rm mix.} = g A^a_\mu \bigl(T^a\phi_0\bigr)_i\,\del^\mu\phi_i~.
\label{lmix}
\ee\vs\nin
Thus, we see that if the associated generator is broken,
i.e. (\ref{notgood}) is satisfied, then there is mixing of the
corresponding gauge boson with the massless $\phi_i$ fields, the
NGBs. It is clear that, just as in the abelian case,  we can eliminate
this term by a suitable gauge transformation on $A^a_\mu$. This would
still leave the mass term unchanged, but would  completely eliminate
the NGBs mixing in (\ref{lmix}) from the spectrum. But even if we
leave the NGBs in the spectrum, and we still have to deal with the
mixing term (\ref{lmix}), we can still see that the gauge boson two
point function remains transverse, a sign that gauge invariance is
still respected despite the appearance of a gauge boson mass. This is
depicted in Figure~\ref{fig:201}.
\vs
\nin
In order to obtain diagram $(c)$ we need to derive the Feynman rule
resulting from the mixing term $\cL_{\rm mix}$ (\ref{lmix}). In
momentum space this becomes
\begin{wrapfigure}[2]{l}{0.3\textwidth}
  \hspace{50pt}\includegraphics[width=.2\textwidth]{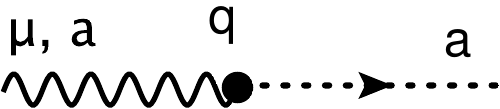}
 \end{wrapfigure}
\vspace{10pt}
\begin{equation}
=  g \bigl(T^a \phi_0\bigr)_i\,q^\mu~,
\end{equation}
\vs\vs\vs\vs
\nin
where the NGB momentum is flowing out of the vertex (its sign changes
if it is flowing into the vertex). The contributions to diagram $(a)$
are transverse as they come from either the leading order propagator
or the loop corrections to it, both already shown to be transverse.
Then the two point function for the gauge boson in the presence of
spontaneous symmetry breaking is
\vs\bear
\Pi_{\mu\nu} &=& \Pi^{(a)}_{\mu\nu} + i M^2_{ab} \,g_{\mu\nu} + g
\bigl(T^a\phi_0\bigr)_i\,q_\mu\,\frac{i\delta_{ab}}{q^2}\, g
\bigl(T^b\phi_0\bigr)_i\,(-q_\nu\bigr)~\nonumber\\
&=& \Pi^{(a)}_{\mu\nu} + i M^2_{ab}\bigl(g_{\mu\nu} -\frac{q_\mu\,a_\nu}{q^2}\bigr)~,
\label{istransverse}
\eear\vs\nin
where to obtain the second line we used (\ref{massmatrixdef}). Then,
just as we saw for the abelian case, we see that the gauge boson two
point function is transverse even in the presence of gauge boson masses.

\subsection{ Example 1: $SU(2)$}

In this first example we gauge the $SU(2)$ of the first example in the
previous lecture.  The lagrangian
\vs\be
\cL = \bigl(D_\mu\phi\bigr)^\dagger\,D^\mu\phi -V(\phi^\dagger\phi)
-\frac{1}{4}F^a_{\mu\nu} F^{a\mu\nu}~,
\label{lagsu2}
\ee\vs\nin
with the covariant derivative on the scalar field is\footnote{We have
  gone back to complex scalar fields for the remaining of the lecture.}
\vs\be
D_\mu\phi(x) =\bigl(\del_\mu - ig A^a_\mu(x) t^a\bigr)\,\phi(x)~,
\label{covder2}
\ee\vs\nin
where the $SU(2)$ generators are given in terms of the Pauli matrices as
\vs\be
t^a=\frac{\sigma^a}{2}~,
\ee\vs\nin
with $a=1,2,3$. Since they transform according to
\vs\be
\phi(x)_j\to e^{i\alpha^a(x) t^a_{jk}}\,\phi_k(x)~,
\ee\vs\nin
with $j,k=1,2$, 
then that are {\em doublets} of $SU(2)$. Since each of the $\phi_j(x)$
are complex scalar fields, we have {\em four} real scalar degrees of
freedom.
We will consider the vacuum
\vs\be
\langle\phi\rangle = \left(\ba{c} 0\\
  \frac{v}{\sqrt{2}}\ea\right)~,
\label{su2vac}
\ee\vs\nin
such that, as required by imposing a non trivial minimum, we have
\vs\be
\langle\phi^\dagger\phi\rangle =\frac{v^2}{2}~,
\label{phi2vac}
\ee\vs\nin
where the factor of $2$ above is chosen for convenience. We are
particularly interested in the gauge boson mass terms. These can be
readily obtained by substituting the vacuum value of the field in the
kinetic term. This is
\vs\bear
\cL_{\rm m}&=&
\bigl(D_\mu\langle\phi\rangle\bigr)^\dagger\,D^\mu\langle\phi\rangle~,\nonumber\\
&=& \frac{g^2}{2}\,A^a_\mu A^{b\mu}\,\bigl(0\quad v\bigr)\,t^a t^b \left(\ba{c}0\\v\ea\right)~, 
\label{gaugemass2}
\eear\vs\nin
 where we used (\ref{su2vac}) in the second line. But for the case of
 $SU(2)$ we can use the fact that
 \vs\be
\bigl\{\sigma^a,\sigma^b\bigr\} = 2\delta^{ab}~,
\ee\vs\nin
which translates into
\vs\be
\bigl\{t^a,t^b\bigr\} = \frac{1}{2} \delta^{ab}~,.
\label{su2acom}
\ee\vs\nin
Then, if we write
\vs\bear
A^a_\mu \,A^{b\mu} \,t^a\, t^b &=& \frac{1}{2} \,A^a_\mu \,A^{b\mu} \,t^a\, t^b  +
\frac{1}{2} \,A^b_\mu\, A^{a\mu} \,t^b \,t^a \nonumber\\
&=& \frac{1}{2} \, A^a_\mu\,A^{b\mu} \, \bigl\{t^a,t^b\bigr\}=
\frac{1}{4}\,A^a_\mu\,A^{a\mu}~,
\label{su2trick}
\eear\vs\nin
where in the last euality we used (\ref{su2acom}). Then we obtain
\vs\be
\cL_{\rm m} = \frac{1}{8} g^2\,v^2\,A^a_\mu\,A^{a\mu}~,
\label{lmass2}
\ee\vs\nin
which results in a gauge boson mass of
\vs\be
M_A = \frac{g\,v}{2}~.
\ee\vs\nin
Notice that {\em all three} gauge bosons obtain this same mass. It is
interesting to compare this result with what we obtained in the
previous lecture for the spontaneous breaking of a {\em global }
$SU(2)$ symmetry using the same vacuum as in (\ref{su2vac}). 
In that case, we saw that all generators were broken, i.e. there are
three massless NGBs in the spectrum and the $SU(2)$ is completely
(spontaneously) broken in the sense that none of its generators leaves
the vacuum invariant. In the case here, where the $SU(2)$ symmetry is
gauged, we see that  all three gauge bosons get masses. This is in
fact the same phenomenon: none of the gauge symmetry leaves the
$SU(2)$ vacuum (\ref{su2vac}) invariant. However, the end result is
three massive gauge bosons, not three massless NGBs. We argued in our
general considerations above that, just as for  the abelian case
before, the NGBs can be removed by a gauge transformation. Let us see
how this can be implemented.
\nin
We consider the following parametrization of the $SU(2)$ doublet scalar field:
\vs\be
\phi(x) = e^{i\pi^a(x) t^a/v}\,\left(\ba{c}0\\
  ~\\
  \frac{v+\sigma(x)}{\sqrt{2}}\ea\right)~,
\label{phiparam2}
\ee\vs\nin
where $\sigma(x)$ and $\pi^a(x)$ with $a=1,2,3$ are real sclalar
fields satisfying
\vs\be
\langle\sigma(x)\rangle=0=\langle\pi^a(x)\rangle~,
\ee\vs\nin
so that this choice of parametrization is consistent with the vacuum
(\ref{su2vac}). Clearly, the potential will not depend on the
$\pi^a(x)$ fields
\vs\be
V(\phi^\dagger\phi)= -\frac{m^2}{2}\,\phi^\dagger\phi
+\frac{\lambda}{2}\,\bigl(\phi^\dagger\phi\bigr)^2~,
\label{pot3}
\ee\vs\nin
The minimization results in\footnote{Notice the different factor in
  the denominator of the second term. This is due to the factor of
  $\sqrt{2}$ in the definition of the vacuum.} 
\vs\be
\langle\phi^\dagger\phi\rangle = \frac{m^2}{2\lambda} ~,
\ee\vs\nin
which results in
\vs\be
v^2=\frac{m^2}{\lambda}~.
\ee\vs\nin
Replacing this in the potential (\ref{pot3}) we obtain
\vs\be
m_\sigma=\sqrt{2\,\lambda}\,v~.
\ee\vs\nin
And of course, the implicit result of having
\vs\be
m_{\pi^1}=m_{\pi^2}=m_{\pi^3}=0~.
\ee\vs\nin
But how do we get rid of the massless NGBs ? If we define the
following gauge transformation
\vs\be
U(x)\equiv e^{-i\pi^a(x) t^a/v}
\ee\vs\nin
under which the fields transform as
\vs\bear
\phi(x) &\to &  \phi'(x) = U(x) \,\phi(x) = \left(\ba{c}0\\
  ~\\
  \frac{v+\sigma(x)}{\sqrt{2}}\ea\right)~,\nonumber\\
~\label{ugauge1}\\
A_\mu &\to & A'_\mu = U(x)\,A_\mu \,U^{-1}(x)
-\frac{i}{g}\,\bigl(\del_\mu U(x)\bigr)\,U^{-1}(x)~,\nonumber
\eear\vs\nin
where we used the notation $A_\mu=A^a_\mu t^a$. It is clear from the
first transformation above, that $\phi'(x)$ does not depend on the
$\pi^a(x)$ fields. Thus, the gauge transformation (\ref{ugauge1}) has
removed them from the spectrum completely. However, the number of
degrees of reedom is the same in boths gauges. We had three transverse
gauge bosons (i.e. 6 degrees of freedom)  and four real scalar fields. In this new gauge we have
three massive gauge bosons (i.e. 9 degrees of freedom) plus one real
scalar, $\sigma(x)$. The total number of degrees of freedom is always
the samee. The gauge were the NGBs diissapear of the spectrum is
called the {\em unitary gauge}.

\subsection{Example 2: $SU(3)$}

We now consider the caso with $G=SU(3)$, and the vacuum is chosen to
be
\vs\be
\langle\phi\rangle =\left(\ba{c}0\\
  0\\\frac{v}{\sqrt{2}}\ea\right)~.
\label{su3vac}
\ee\vs\nin
The $SU(3)$ generators $t^a$, with $a=1,\cdots,8$,  where explicitely written in the previous
lecture. They generaly satisy the group algebra and normalization
\vs\bear
[t^a,t^b] &=& i f^{abc} t^c\nonumber\\
\Tr\bigl[t^a\,t^b\bigr] &=& \frac{1}{2}\,\delta^{ab}~.\nonumber
\eear\vs\nin
In addition to these, we can write the anticommutator as
\vs\be
\bigl\{t^a,t^b\bigr\} = \frac{1}{3} \delta^{ab} + 
d^{abc}\,t^c~,
\label{su3acom}
\ee\vs\nin
where the $d^{abc}$ are totally {\em symmetric} constants\footnote{In
  general, this expression is valid for $SU(N)$ groups, with the $3$
  replaced by $N$. For the particular case of $N=2$ the $d^{abc}$
  vanish.} . We want to compute the gauge boson masses. Just as we
did for the $SU(2)$ case in (\ref{gaugemass2}), the gauge boson mass terms come from the
kinetic terms with $\phi$ replaced by the vacuum (\ref{su3vac}).
\vs\bear
\cL_{\rm m} &=& \bigl(D_\mu \langle\phi\rangle\bigr)^\dagger D^\mu\langle\phi\rangle
= \frac{g^2}{2}\,A^a_\mu\,A^{b\mu}\,\bigl(0\quad 0\quad
v\bigr)\,t^a\,t^b\,\left(\ba{c}0\\
  0\\ v\ea\right)\nonumber\\
&=& \frac{g^2}{2}\,A^a_\mu\,A^{b\mu}\,\bigl(0\quad 0\quad
v\bigr)\,\frac{1}{2} \,\bigl\{t^a,t^b\bigr\}\, \left(\ba{c}0\\
  0\\ v\ea\right)\label{su3gmass1}
\eear\vs\nin
wher in the second line we used the same trick as in $SU(2)$ to write
the anticommutator. Then, using (\ref{su3acom}) we obtain
\vs\be
\cL_{\rm m} = \frac{g^2}{4}\,\,A^a_\mu\,A^{b\mu}\,\bigl(0\quad 0\quad
v\bigr)\,\Bigl\{\frac{1}{3}\,\delta^{ab} + d^{abc}\,t^c\Bigr\}\, \left(\ba{c}0\\
  0\\ v\ea\right)\label{su3gmass2}~.
\ee\vs\nin
In order to proceed further in the computation if the gauge boson
masses we need to know the values of the $d^{abc}$ constants for
$SU(3)$. The non zero values are
\vs\bear
d^{118}=d^{228}=d^{338}=-d^{888}=\frac{1}{\sqrt{3}}\nonumber\\
d^{448}=d^{558}=d^{668}=d^{778}= -\frac{1}{2\sqrt{3}}\nonumber\\
d^{146}=d^{157}=-d^{247}=d^{256}=d^{344}= d^{355}
=-d^{366}=-d^{377}=\frac{1}{2}~.
\label{dvalues}
\eear\vs\nin
However,  when it comes to the masses of the gauge bosons for
$a,b=1,2,3$ there is a simpler way to compute them.
We know from Section~\ref{sec3}, equation (\ref{su3pauli})  that these geerators can be written
as
\vs\be
t^i=\frac{1}{2}\left(\ba{cc}
  \sigma^i&0\\
  0\,\,\,0&0\ea\right)~,
\nonumber\ee\vs\nin
with the $\sigma^i$ the Pauli matrices for $i=1,2,3$. We also know
that they annihilate the vacuum defined in (\ref{su3vac}), that is
\vs\be
t^i\,\langle\phi\rangle =0~,
\ee\vs\nin
for $i=1,2,3$. 
Thus, just by looking at
the first line in (\ref{su3gmass1}) we can see that the gauge bosons
$A^1_\mu$, $A^2_\mu$ and $A^3_\mu$ will be massless, that
is\footnote{It is clear that all the elements of the $3\times3$ mass
  matrix will vanish, both diagonal and non diagonal.} 
\vs\be
M_{A^1}=M_{A^2}=M_{A^3}=0~.
 \ee\vs\nin
Once again, it is interesting to compare with the spontaneous breaking
of the global $SU(3)$ symmetry studied in the previous lecture. There
we saw that the first three generators of $SU(3)$ still left the vacuum
invariant, and therefore there were no massless NBGs associated with
them. Here, the fact that the $SU(2)$ subgroup of $SU(3)$ defined by
the generators in (\ref{su3pauli}) still preserve the vacuum
(\ref{su3vac}) results in those gauge bosons remaining massless. It is
in this sense that the $SU(2)$ remains {\em unbroken}.

\nin
Conversely, we expect that the {\em broken } generators, i.e. those
that do not annihilate the vacuum (\ref{su3vac}), will be associated
with massive gauge bosons, as they were associated with the existence
of massless NGBs in the case of the global symmetry. To obtain their
masses we first observe that the {\em off diagonal } mass terms in
(\ref{su3gmass2}), i.e. those for $a\not=b$ are zero, even for
$a,b=4,5,6,7,8$. To see this, we write the corresponding mass terms as
\vs\be
\cL^{a\not=b}_{\rm m} = \frac{g^2}{4}\,\,A^a_\mu\,A^{b\mu}\,\bigl(0\quad 0\quad
v\bigr)\, d^{abc}\,t^c\, \left(\ba{c}0\\
  0\\ v\ea\right)\label{su3gmassnond}~.
\ee\vs\nin
 The terms contributing are those with $d$'s in the last line of
 (\ref{dvalues}). For these, the generators $t^c$ appearing in
 (\ref{su3gmassnond}) correspond to $c=4,5,6$ and $7$. But these
 generators, when acting on the vacuum in (\ref{su3gmassnond}) do not
 leave it invariant. They will result in a column vector that is
 orthogonal to the vacuum in (\ref{su3vac}), i.e. will give zero when
 it hits the row vector $(0\quad 0\quad v)$. Then, all these non
 diagonal mass terms vanish. Finally we consider the diagonal mass
 terms, i.e. $a=b=4,5,6,7,8$. Their contributions come both from the
 $\delta^{ab}$ as well as from the $d$'s in the second line in
 (\ref{dvalues}). Let us explicitely compute the case of $a=b=4$.
 We have
 \vs\be
\cL^{a=b=4} _{\rm m} = \frac{g^2}{4}\,\,A^4_\mu\,A^{4\mu}\,\bigl(0\quad 0\quad
v\bigr)\,\Bigl\{\frac{1}{3}+ d^{448}\,t^8\Bigr\}\, \left(\ba{c}0\\
  0\\ v\ea\right)\label{su3gmassab4}~.
\ee\vs\nin
 When substituting for the values of $d^{448}$ and for $t^8$ (see
   previous lecture) we obtain
   \vs\be
\cL^{a=b=4}_{\rm m}  = \frac{1}{8}\,g^2\,v^2\,A^4_\mu\,A^{4\mu}~,
   \ee\vs\nin
   which results in
   \vs\be
M_{A^4}= \frac{g\,v}{2}~.
\ee\vs\nin
However, since all the $d^{ab8}$ with $a=b\not=8$ have the same value and
only $t^8$ contributes, then it is clear that we will have
\vs\be\boxed{
M_{A^4}=M_{A^5}=M_{A^6}=M_{A^7} = \frac{g\,v}{2}~}.
\ee\vs\nin
Finally, taking into account the different value of $d^{888}$, we
obtain
\vs\be\boxed{
M_{A^8} = \frac{g\,v}{\sqrt{3}}~}.
\ee\vs\nin
These completes the spectrum of gauge bosons: five massive, and 3
massless ones. Just as we expect from the fact that the spontaneous
symmetry breaking induced by the vacuum (\ref{su3vac})
respects $SU(2)$. We say that the spontanous symmetry breaking pattern
is
\vs\be
SU(3)\to SU(2)~.
\ee\vs\nin

\subsection{The Electroweak Standard Model}

\nin
For an example of spontaneous breaking of a gauge symmetry we look no
further than to the standard model (SM) of particle physics \footnote{An expanded and
  more detailed presentation can be found in Ref.~\cite{Burdman:2024uba}}. What is
broadly called the SM is a quantum field theory that describes all the interactions of all the known
elementary particles, with the exception of a quantum description of gravity. These are: the strong interactions, as described by
quantum chromodynamics (QCD), an unbroken $SU(3)$ gauge theory; and the electroweak interactions as
described by the gauge symmetry $SU(2)_L\times U(1)_Y$. It is the
latter that is spontaneously broken to give rise to three massive weak
gauge bosons and a massless photon responsible for the electromagnetic interactions. 
Since the QCD gauge symmetry is not spontaneously broken by the
Anderson-Higgs mechanism, we will ignore it in what follows since  is a
``spectator'' interaction that does not change when the electroweak
symmetry is broken to electromagnetism.

\subsubsection{ Why $SU(2)_L\times U(1)_Y$ ? }  

\nin
The electroweak gauge group is a product of an $SU(2)$ gauge symmetry
and a $U(1)$ one. How do we know this is the electroweak gauge
symmetry group ? And, what is the meaning of the subscripts $L$ and
$Y$ ? Although we will not attempt to go through all the fascinating
history of experimental evidence that goes into building the
electroweak SM,  it is worthwhile to understand the main points that
result in the building of this strange looking, yet amazingly
successful description of fundamental physics.

\nin
Let us review the main evidences leading to the gauge structure of the
electroweak theory.
\begin{itemize}
\item \underline{Weak Interactions (Charged)}: Weak decays, such as
  $\beta$ decays $n\to p\,e^- \bar\nu_e$ or
  $\mu^-\to\nu_\mu\,\bar\nu_e\,e^-$ among many others, are mediated by
  {\em charged} currents. Let us look at the case of muon decay. It is
  very well described by a four fermion interaction, i.e. with a  non
  renormalizable coupling $G_F$, the Fermi constant. In fact, all
  other weak interactions can be described in this way with the same
  Fermi constant ( to a very good approximation, more later). The
  relevant Fermi lagrangian is 
  \vs\be
\cL_{\rm Fermi} = \frac{G_F}{\sqrt{2}}\,\bigl(\bar\mu_L\gamma_\mu
  \nu_L\bigr)\,\bigl(\bar e_L \gamma^\mu \nu_e\bigr)~,
  \label{fermi1}
  \ee\vs\nin
  where we already included the fact that the charged weak interactions
  only involve {\em left handed fermions}. That is, the
  phenomenolgically built Fermi lagrangian above tells us that the
  weak decay of a muon is described by the product of two {\em charged
    vector currents} coupling only left handed fermions. 
  The fact that only left handed
  fermions participate in the charged weak interactions is an
  experimentally established fact, observed in {\em all charged weak
    interactions}. This is done by a variety of experimental
  techniques. Fors instance,  in the case of muon decay, the angular
  distribution of the outgoing electron is very different if this is
  left or right handed. Precise measurements (performed over decades
  of increasingly accurate experiments) have concluded that the
  outgoing electron is left handed only. The different couplings
  involving left  and right handed fermions require {\em parity
    violation}. Moreover, the charged weak interactions require {\em
    maximal parity violation}: only one handedness participate.  
If we  assume that the non renormalizable four fermion interaction is the
  result of integrating out a gauge boson with a renormalizable interaction, this would point to the
  need of $2$ charged gauge bosons.This is schematically shown in
  Figure~\ref{fig:203}. Assuming that $m_\mu \ll M_W$, we integrate
  out the massive vector gauge boson to obtain
  \vs\be
\frac{G_F}{\sqrt{2}} = \frac{g^2}{8 M_W^2}~,
  \ee\vs\nin
  where $g$ is the renormalizable coupling of the gauge bosons to
  fermions in diagram $(b)$.  The charged vector gauge bosons, $W^\pm$ were
  discovered in the 1980s and studied with gerat detail ever since.

 \vs
\begin{figure}[h]
  \begin{center}
    \includegraphics[width=.6\textwidth]{fig_20-3}
\end{center}
\caption{Diagram  $(a)$ is the Feyman diagrams associated with the four
  fermion Fermi lagrangian (\ref{fermi1}). Diagram $(b)$ shows the
  corresponding exchange of a massive charged gauge boson, $W_\mu^\pm$. 
}
\label{fig:203}
\end{figure}
\nin

\item \underline{Weak Neutral Currents}: In addition to the charged
  currents described by (\ref{fermi1}), we have know since
  experimental evidence first appeared in the 1970s, that there are
  also {\em neutral weak currents}. These were first observed by
  neutrino scattering off nucleons. Normally, the charged currents
  would result in $\nu_e\,N \to e^- \,N'$, with $N$ and $N'$ protons
  and neutrons. This is just a crossed diagram of $\beta$ decay. But
  the reaction $\nu\,N\to \nu\,N$ was also observed. Many other
  reactions involving neutral currents have been observed since
  then. They also violate parity. However, they do not do so
  maximally. This menas that the neutral current, or the vector gauge
  boson that we need to integrate out to obtain them at low energies,
  couple differently to lef and right handed fermions but, unlike the
  charged currents, the do couple to right handed fermions.
  The neutral vector gauge boson, $Z^0$, was also discovered in the
  1980s and its properties studied with great precision. 

  \item \underline{Electromagnetism}: Of course, we know that the
    electromagnetic interactions are described by a quantum field
    theory, QED, mediated by a neutral {\em massless} vector gauge
    boson, the photon. One important feature to remember is that the
    photon coupling in QED is {\em parity invariant}. No parity
    violation is present in QED.

  \end{itemize}

  The elements described above suggest that we need: $4$ gauge bosons
  for a unified description of the weak and electromagnetic
  interactions. Three of them are massive, one (the photon) must
  remain massless after spontaneous symmetry breaking. The SM gauge
  group is then $G=SU(2)\times U(1)$ which matches the number of gauge
  bosons. However, we know that two of these only couple to left
  handed fermions. whereas one of the massive ones (the neutral)
  couples differently to left and right handed fermions. Finally, the
  photon must remain massless and its couplings parity invariant. The
  choice of gauge group is then
  \vs\be\boxed{
  G = SU(2)_L\times U(1)_Y~},
  \label{smgroup}
  \ee\vs\nin
  where the three gauge bosons couple to left handed fermions only,
  and the $U(1)_Y$ is not identified with the $U(1)_{\rm EM}$, the
  abelian gauge symmetry responsible for electromagnetism. As we will
  see below, two of the $SU(2)_L$ gauge bosons will result in the
  $W^\pm_\mu$. On the other hand to obtain the $Z^0$ and the photon we
  will need to carefully choose the pattern of spontaneous symmetry
  breaking of the SM gauge group in (\ref{smgroup}) down to $U(1)_{\rm  EM}$.

  \subsubsection{Spontaneous Breaking of $SU(2)_L\times U(1)_Y \to
    U(1)_{\rm EM}$}

  \nin
  All matter in the SM (i..e. fermions and scalars) will have some
  transformation property under the SM gauge group $G$. This means
  that we need to assign to each particle a charge under the abelian
  group factor, the $U(1)_Y$. This is called {\em hypercharge}, since
  it is not quite the electric charge.\\
  \nin
  We first consider a scalar field $\Phi$ in the fundamental representation
  of $SU(2)$ and with hypercharge $U(1)_Y$,
  \vs\be
  Y_{\Phi}=1/2~.
  \ee\vs\nin
  The scalar doublet can be written as
  \vs\be
  \Phi = \left(\ba{c}\phi^+\\\phi^0\ea\right)~,
  \label{hdoublet}
  \ee\vs\nin
  where $\phi ^+$ and $\phi^0$ are complex scalar fields, resulting in
  four real scalar degrees of freedom\footnote{At this point, the
    labels ``$+$'' and ``$0$'' are just arbitrary, since we have not
    even defined electric charges But these label will be consistent in
    the future, after we have done this.}.    Under a $SU(2)_L\times
  U(1)_Y$ gauge transformation, the Higgs doublet transforms as
  \vs\be
  \Phi(x)\to e^{i \alpha^a(x) t^a}\,e^{i \beta(x) Y_{\Phi}},\Phi(x)~,
  \label{hdtransf}
  \ee\vs\nin
  where $t^a$ are the $SU(2)_L$ generators (i.e. Pauli matrices
  divided by $2$), $\alpha^a(x)$ are the three
  $SU(2)_L$  gauge
  parameters, $\beta(x)$ is the $U(1)_Y$ gauge parameter, and it is
  understood that the $U(1)_Y$ factor of the gauge transformation
  contains a factor of the identity $I_{2\times 2}$ after the
  hypercharge $Y_{\Phi}$. Thus, the covariant derivative on $\Phi$ is
  given by
  \vs\be
D_\mu\Phi(x) = \Bigl(\del_\mu -ig A^a_\mu(x) t^a -ig' B_\mu(x)
Y_{\Phi} I_{2\times 2}\Bigr)\,\Phi(x)~.
\label{covderphi}
\ee\vs\nin
Here, $A^a_\mu(x)$ is the $SU(2)_L$ gauge boson, $B_\mu(x)$ the
$U(1)_Y$ gauge boson, and $g$ and $g'$ are their corresponding
couplings. The lagrangian of  the scalar and gauge sectors of the SM is then
\vs\be
\cL = \bigl(D_\mu\Phi\bigr)^\dagger D^\mu\Phi - V(\Phi^\dagger \Phi)
-\frac{1}{4} F^a_{\mu\nu} F^{a\mu\nu} -\frac{1}{4} B_{\mu\nu}
B^{\mu\nu}~,
\label{smlag1}
\ee\vs\nin
where $F^a_{\mu\nu}$ is the usual $SU(2)$ field strength built out of
the gauge fields $A^a_\mu(x)$ and $B_{\mu\nu}$ is the $U(1)_Y$ field
strength given by the abelian expression
\vs\be
B_{\mu\nu} = \del_\mu B_\nu(x) -\del_\nu B_\mu(x) ~.
\label{bmunu}
\ee\vs\nin
As usual, we consider the potential
\vs\be
V(\Phi^\dagger\Phi) = -m^2\bigl(\Phi^\dagger\Phi\bigr) + \lambda
\bigl(\Phi^\dagger\Phi\bigr)^2~,
\label{smpotential}
\ee\vs\nin
which is minimized for
\vs\be
\langle\Phi^\dagger\Phi\rangle = \frac{m^2}{2\lambda}\equiv
\frac{v^2}{2}~.
\label{vmindef}
\ee\vs\nin
In order to fulfil this, we choose the vacuum
\vs\be
\langle\Phi\rangle = \left(\ba{c}0\\
  \frac{v}{\sqrt{2}}\ea\right)~.
\label{smvac}
\ee\vs\nin
Just as in the previous examples of SSB of non abelian gauge
symmetries, the next question is what is the symmetry breaking
pattern, i.e. what gauge bosons get what masses, if any. In
particular, we want one of the four gauge bosons in $G$ to remain
massless after imposing the vacuum $\langle\Phi\rangle$ in
(\ref{smvac}). This means that there must be a generator or, in this
case, a linear combination of generators of $G$ that annihilates
$\langle\Phi\rangle$, leaving the vacuum invariant under a $G$
transformation. This combination of generators must be associated with
the massless photon in $U(1)_{\rm EM}$, the remnant gauge group after
the spontaneous breaking. One trick to identify this combination of
generators is to consider the gauge transformation defined by
\vs\bear
\alpha^1(x)=\alpha^2(x)=0\nonumber\\
\alpha^3(x)=\beta(x)~.
\label{t3plusygauge}
\eear\vs\nin
The exponent in the gauge transformation has the form
\vs\bear
i\alpha^3(x) t^3 + i \beta(x) Y_\Phi I_{2\times2} &=& i
\frac{\beta(x)}{2} \Bigl[\left(\ba{cc}1&0\\0&-1\ea\right) +
\left(\ba{cc}1&0\\0&1\ea\right)\Bigr]\nonumber\\
&=&\frac{i\beta(x)}{2}\left(\ba{cc}1&0\\0&0\ea\right)~.
\label{charge1}
\eear\vs\nin
Then we see that this combination
\vs\be\boxed{
\bigl(t^3+Y_\Phi\bigr)\,\langle\Phi\rangle =0~},
\label{killsvac}
\ee\vs\nin
indeed annihilates the vacuum, leaving it invariant. 
Thus, we suspect that this linear combination of $SU(2)_L\times
U(1)_Y$ generators must be associated with the massless photon. We
will come back to this point later.
\\
\nin
We now go to extract the gauge boson mass terms from the scalar
kinetic term in (\ref{smlag1}). This is
\vs\bear
\cL_{\rm m} &=& \bigl(D_\mu\langle\Phi\rangle\bigr)^\dagger
D^\mu\langle\Phi\rangle\nonumber\\
&=& \frac{1}{2}\,(0\quad v)\,\bigl(g A^a_\mu t^a
+ g' Y_\Phi B_\mu \bigr) \bigl(g A^{b\mu} t^b
+ g' Y_\Phi B^\mu \bigr)\,\left(\ba{c}0\\v\ea\right)~.
\label{gbmass2}
\eear\vs\nin
For the product of the two $SU(2)$ factors we will use the trick in
(\ref{su2trick}). Then, the only terms we need to be careful about are
the mixed ones: one $SU(2)$ times one $U(1)_Y$ contribution. There are
two of them, and each has the form
\vs\be
\frac{1}{2}\,(0\quad v)\, g\,g' \frac{\sigma^3}{2}
Y_\Phi\,\left(\ba{c}0\\v\ea\right)
= -\frac{1}{2} \frac{v^2}{4}\,g\,g'
\,A^3_\mu\,B^\mu~,
\label{su2u1mix}
\ee\vs\nin
where in the second equality we used $Y_\Phi=1/2$.
We then have
\vs\be
\cL_{\rm m} = \frac{1}{2}\,\frac{v^2}{4}\Bigl\{g^2A^1_\mu A^{1\mu}+g^2
A^2_\mu A^{2\mu} +g^2A^3_\mu A^{3\mu} + g'^2 B_\mu B^\mu -2g g'
A^3_\mu B^\mu\Bigr\}~.
\label{gbmass3}
\ee\vs\nin
From this expression we can clearly ee that $A^1_\mu$ and $A^2_\mu$
acquire masses just as we saw in the pure $SU(2)$ example. It will be
later  convenient  to define the linear combinations
\vs\be
W^\pm_\mu\equiv \frac{A^1_\mu\mp iA^2_\mu}{\sqrt{2}}~,
\label{wpmdef}
\ee\vs\nin
which allows us to write the first two terms in (\ref{gbmass3}) as
\vs\be
\cL_{\rm m}^W = \frac{g^2\,v^2}{4}\,W^+_\mu W^{-\mu}~.
\label{wmassterm}
\ee\vs\nin
These two states have masses
\vs\be
M_W=\frac{g\,v}{2}~.
\label{wmass1}
\ee\vs\nin
On the other hand, the
fact that $A^3_\mu$ and $B_\mu$ have a mixing term prevents us from
reading off masses. We need to rotate these states to go to a bases
without mixing, a diagonal basis. In order to clarify what needs to be
done, we can write the last three terms in (\ref{gbmass3} in matrix
form
\vs\be
\cL_{\rm m}^{\rm neutral} = \frac{1}{2}\frac{v^2}{4}
(A^3_\mu\quad B_\mu)\,\left(\ba{cc} g^2&-g\,g'\\
  -g\,g'&g'^2\ea\right) \,\left(\ba{c}A^{3\mu}\\B^\mu\ea\right)~,
\label{neutralmass1}
\ee\vs\nin
where the task is to find the eigenvalues and eigenstates of the
matrix above. It is clear that one of the eigenvalues is zero, since
the determinant vanishes. Then the squared masses of the physical neutral
gauge bosons are
\vs\bear
M^2_\gamma&=&0\nonumber\\
~\label{neutralmass2}\\
M^2_Z &=&\frac{v^2}{4}\,(g^2+g'^2)~\nonumber 
\eear\vs\nin
The eigenstates in terms of $A^3_\mu$ and $B_\mu$, the original $SU(2)_L$ and
$U(1)_Y$ gauge bosons respectively, are
\vs\bear\boxed{
A_\mu \equiv \frac{1}{\sqrt{g^2+g'^2}}\,\bigl(g' A^3_\mu + g
B_\mu\bigr)~}\label{photondef1}\\
~\nonumber\\
\boxed{Z_\mu\equiv \frac{1}{\sqrt{g^2+g'^2}}\,\bigl(g A^3_\mu - g'
B_\mu\bigr)~}.\label{zdef1}
\eear\vs\nin
Alternatively, we could have obtained the same result by defining an
orthogonal rotation matrix to diagonalize the interactions above.
That is,  rotating the states by
\vs\be
\left(\ba{c}Z_\mu\\A_\mu\ea\right)=
\left(\ba{cc}\cos\theta_W & -\sin\theta_W\\
  \sin\theta_W&\cos\theta_W\ea\right)\,\left(\ba{c}
  A^3_\mu\\B_\mu\ea\right)~,
\label{wangledef1}
\ee\vs\nin
results in diagonal neutral interactions if we have
\vs\be
\cos\theta_W\equiv \frac{g}{\sqrt{g^2+g'^2}},\qquad
\sin\theta_W\equiv \frac{g'}{\sqrt{g^2+g'^2}},
\label{wangledf2}
\ee\vs\nin
where $\theta_W$ is called the Weinberg angle. It is useful to
invert (\ref{wangledef1}) to obtain
\vs\bear
A^3_\mu &=& \sin\theta_W A_\mu +\cos\theta_W Z_\mu\label{a3asaz}\\
~\nonumber\\
B_\mu &=& \cos\theta_W A_\mu -\sin\theta_W Z_\mu~.\label{basaz}
\eear\vs\nin
Using these expressions for $A^3_\mu$ and $B_\mu$ we can replace them
in the covariant derivative acting on the scalar doublet $\Phi$. Their
contribution fo $D_\mu$ is
\vs\bear
-i g A^3_\mu t^3 - i g' Y_\Phi B_\mu &=&
-iA_\mu\bigl(g\sin\theta_Wt^3 +  g'\cos\theta_W
Y_\Phi\bigr) 
- i\bigl(g\cos\theta_W t^3 - g'\sin\theta_W Y_\Phi\bigr)\,Z_\mu
\nonumber\\
~\label{photonandzcpl}\\
&=&- i g\sin\theta_W \Bigl(t^3+Y_\Phi\Bigr) A_\mu 
-i\frac{g}{\cos\theta_W}\Bigl(t^3 -(t^3+Y_\Phi)\sin^2\theta_W\Bigr) Z_\mu~,
\nonumber
\eear\vs\nin
where it is always understood that the hypercharge $Y_\Phi$ is always
multiplied by the identity, and in the last identity we used the fact
that
\vs\be
g' \cos\theta_W = g \sin\theta_W~,
\ee\vs\nin
and trigonometric identities. We can conclude that is $A_\mu$  is to
be identified with the photon field, then its coupling must be $e$
times the charged of the particle it is coupling to (e.g. $-1$ for an
electron.
Thus we must impose that
\vs\be\boxed{
  e= g\,\sin\theta_W~},
\label{qedcdef1}
\ee\vs\nin
and that the charge operator,  acting here on the field $\Phi$ coupled to $A_\mu$ is
defined as
\vs\be\boxed{
Q = t^3 + Y_\Phi~}.
\label{emchargedef}
\ee\vs\nin
Then we can read the photon coupling to the doublet scalar field $\Phi$ from
\vs\be
-i\, e\, A_\mu \,Q \,\Phi(x) = -i \,e \,A_\mu \,Q
\,\left(\ba{c}\phi^+\\\phi^0\ea\right) ~.
\ee\vs\nin
Substituting $Y_\Phi=1/2$ we have
\vs\be
Q \,\left(\ba{c}\phi^+\\\phi^0\ea\right) = \left(\ba{cc}1&0\\
  0&0\ea\right) \,\left(\ba{c}\phi^+\\\phi^0\ea\right) = \,\left(\ba{c}\phi^+\\0\ea\right) ~,
\ee\vs\nin
which tells us that the top complex field in the scalar doublet has
charge equal to $1$ (in units of $e$, the proton charge), whereas the
bottom component has zero charge, justifying our choice of labels.
On the other hand, we see that fixing $Q$ to be the electromagnetic
charge operator, completely fixes the couplings of $Z_\mu$ to the
scalar $\Phi$. This is now,  from (\ref{photonandzcpl}),
\vs\be
-i \frac{g}{\cos\theta_W} Z_\mu \Bigl(t^3
-Q\sin^2\theta_W\Bigr)\,\Phi~.
\label{zcpltophi}
\ee\vs\nin
We will see below that the choice of fixing the $A_\mu$ couplings to
be those of electromagnetism, fixes completely the $Z_\mu$ to all
elementary couplings, giving a wealth of predictions.

\subsubsection{Gauge Couplings of Fermions}

\nin
The SM is a {\em chiral gauge theory}, i.e. its gauge couplings differ
for different chiralities.
To extract the left handed fermion gauge couplings, we look at the covariant derivative
\vs\be
D_\mu\psi_L = \bigl(\del_\mu -igA^a_\mu t^a -i g' Y_{\psi_L} B_\mu
\bigr) \psi_L~,
\label{lhfcd1}
\ee\vs\nin
where $Y_{\psi_L}$ is the left handed fermion hypercharge. On the
other hand, since right handed fermions do not feel the $SU(2)_L$
interaction, their covariant derivative is given by
\vs\be
D_\mu\psi_R = \bigl(\del_\mu -i g' Y_{\psi_R} B_\mu
\bigr) \psi_R~,
\label{rhfcd1}
\ee\vs\nin
with $Y_{\psi_R}$ its hypercharge.
Using the covariant derivatives above, we can extract the neutral and
charged couplings. We start with the neutral couplings, which in terms
of the gauge boson mass eigenstates are the couplings to the photon
and the $Z$.

\vs\vs\nin
\underline{Neutral Couplings}: 
From (\ref{lhfcd1}), the neutral gauge couplings of a left handed fermions
are 
\vs\bear
\bigl(-ig t^3 A^3_\mu-ig' Y_{\psi_L} B_\mu\bigr)\psi_L &=&
ig\sin\theta_W\bigl(t^3+Y_{\psi_L}\bigr) \,A_\mu\,\psi_L \nonumber\\
&-&i\frac{\bigl(g^2 t^3-i g'^2 Y_{\psi_L}\bigr)}{\sqrt{g^2+g'^2}} \,Z_\mu \,\psi_L~,
  \label{neutralf1}
  \eear\vs\nin
  where on the right hand side we made use of (\ref{a3asaz}) and
  (\ref{basaz}).
  Now, we know that the photon coupling should be
  \vs\be
-ie \,Q_{\psi_L}~,
\ee\vs\nin
with $Q_{\psi_L}$ the fermion electric charge operator. Thus, we must
identify
\vs\be
Q_{\psi_L} = t^3+ Y_{\psi_L}~,
\label{fermioncharge1}
\ee\vs\nin
as the fermion charge. We can use our knowledge of the fermion charges
to fix their hypercharges. As an example, let us consider the left
handed lepton doublet. For the lightest family, this is written in the notation 
\vs\be
L = \left(\ba{c}\nu_{eL}\\e^-_L\ea\right)~.
\label{lepdoublet1}
\ee\vs\nin
The action of $t^3$ on $L$ is
\vs\bear
t^3\,L&=& \left(\ba{cc}1/2&0\\
  0&-1/2\ea\right)\, \left(\ba{c}\nu_{eL}\\e^-_L\ea\right)\nonumber\\
~\nonumber\\
&=&\left(\ba{c}(1/2)\,\nu_{e_L}\\(-1/2)\, e^-_L\ea\right) \equiv \left(\ba{c}
  t^3_{\nu_{e_L}}\,\nu_{e_L}\\t^3_{e_L}\, e^-_L\ea\right)~,
\label{t3onl}
\eear\vs\nin
where in the last equality we defined $t^3_{\nu_{e_L}}=1/2$ and
$t^3_{e_L}$ as the eigenvalues of the operator $t^3$ associated to the
electron neutrino and the electron. Then, we have
\vs\be
Q_{L}\,L = \left(\ba{cc} 1/2+Y_L&0\\
  -1/2+Y_L &0\ea\right)\, \left(\ba{c}\nu_{eL}\\e^-_L\ea\right) =
\left(\ba{c}(1/2+Y_L)\,\nu_{e_L} \\
  (-1/2+Y_L) \, e_L\ea\right)~.
\label{qonl} 
\ee\vs\nin
But we know that the eigenvalue of the charge operator applied to the
neutrino must be zero, as well as that the eigenvalue of the electron
must be $-1$. Thus, we obtain the hypercharge of the left handed lepton
doublet
\vs\be\boxed{
  Y_L=-\frac{1}{2}~},
  \label{lhlddhyp}
\ee\vs\nin
which is fixed to give us the correct electric charges for the members
of the doublet $L$.
\nin
We can do the same with the right handed fermions. These, however do
not have $t^3$ in the covariant derivative (see (\ref{rhfcd1})
). Then, for $e^{-}_R$, the right handed electron, we have that
$t^3_{e_R}=0$, which means that, since 
\vs\be
Q_{e^-_R}=  -1~,
\ee\vs\nin
then the right handed electron's hypercharge is equal  to it:
\vs\be\boxed{
Y_{e^-_R} =-1~}.
\ee\vs\nin
Similarly, the right handed electron neutrino has zero electric
charge, which results in
\vs\be\boxed{
Y_{\nu_R}=0~} .
\ee\vs\nin
Now that we fixed all the lepton hypercharges by imposing that they
have the QED couplings to the photon, we can extract their couplings
to the $Z$ as predictions of the electroweak SM.  From
(\ref{neutralf1}) we have
\vs\bear
-i \bigl(g\,\cos\theta_W \,t^3 - g'\,\sin\theta_W \,Y_\psi\bigr) Z_\mu\,
\psi
&=& -i \frac{g}{\cos\theta_W}\,\bigl( \cos^2\theta_W\,t^3 -
\sin^2\theta_W\,Y_\psi\bigr)\,Z_\mu\,\psi\nonumber\\
~\nonumber\\
&=& -i\,\frac{g}{\cos\theta_W} \,\bigl(t^3 -
\sin^2\theta_W\,Q_\psi\bigr)\,Z_\mu\,\psi~,
\label{neutralfcoup2}
\eear\vs\nin
where the initial expressios makes use of $\cos\theta_W $ and
$\sin\theta_W$ in terms of $g$ and $g'$, in the first equality we used
that $\tan\theta_W=g'/g$ and, in the final equality, we  used that in
general $Q_\psi=t^2+Y_\psi$, independently of the fermion chirality, as
long as we generalize (\ref{fermioncharge1}) for right handed fermions
using $t^3_{\psi_R}=0$. For instance, from (\ref{neutralfcoup2}) we can read off the
lepton couplings of the $Z$ boson. These are,
\vs\bear
\nu_{e_L} : &&\qquad
-i\frac{g}{\cos\theta_W}\,\Bigl(\frac{1}{2}\Bigr)\nonumber\\
  e^-_L: &&\qquad  -i\frac{g}{\cos\theta_W}\,\Bigl(-\frac{1}{2}
  +\sin^2\theta_W\Bigr)\nonumber\\
  ~\label{lzcoup1}\\
  e^-_R: &&\qquad -i\frac{g}{\cos\theta_W}\,\Bigl(
  -\sin^2\theta_W\Bigr)\nonumber\\
  \nu_{e_R}: &&\qquad 0\nonumber~.
\eear\vs\nin
From the couplings above, we see that every lepton has a different
predicted coupling to the $Z$. These are, of course, three level
predictions. Measurements of these $Z$ couplings have been performed
with subpercent precision for a long time, and the SM predictions for
the fermion gauge couplings have
passed the tests every time. Another, interesting point, is that right
handed neutrinos have {\em no gauge couplings} in the SM: no $Z$
coupling, certainly no electric charge and no QCD couplings. Thus,
from the point of view of the SM, the right handed neutrino need not
exist.

\subsubsection{Charged Couplings}:

\nin
We complete here the derivation of the gauge couplings of leptons by extracting their
charged couplings.These come from the $SU(2)_L$ gauge couplings, as we
see from
\vs\be
-i g\bigl(A^1_\mu t^1 + A^2_\mu t^2\bigr) =
-i\frac{g}{\sqrt{2}}\,\left(\ba{cc}0& W^+_\mu\\
  W^-_\mu&0\ea\right)~,
\label{chargedgauge2}
\ee\vs\nin
which then involve only left handed fermions. 
Then, from the gauge part of the left handed doublet kinetic term 
\vs\be
\cL_{L} = \bar L i\dslcv D L ~,
\ee\vs\nin
we obtain their charged couplings 
\vs\bear
\cL_L^{\rm ch.} &=& (\bar\nu_{e_L}\quad \bar e_L)\,\gamma^\mu
\frac{g}{\sqrt{2}}\,
\left(\ba{cc}0& W^+_\mu\\
  W^-_\mu&0\ea\right)\,\left(\ba{c}\nu_{e_L}\\e_L\ea\right)\nonumber\\
~\label{chargedcoup2}\\
&=& \frac{g}{\sqrt{2}}\,\Bigl\{\bar\nu_{e_L}\gamma^\mu e_L \, W^+_\mu
+ \bar e_L\gamma^\mu \nu_{e_L}\,W^-_\mu\Bigr\}~,
\nonumber
\eear\vs\nin
where we can see that, as required by hermicity,  the second term is the hermitian conjugate of
the first. The Fermi lagrangian can be obtained from $\cL_L^{\rm ch.}$
by integrating out the $W^\pm$ gauge bosons.

\nin
We now briefly comment on the electroweak gauge coulpings of quarks.
Just as for leptons, we concentrate on the first family. The left
handed quark doublet is
\vs\be
q_L = \left(\ba{c}u_L\\d_L\ea\right)~,
\label{qldef1}
\ee\vs\nin
We know  that, independently of helicity, the charges of the up and
down quarks are $Q_u=+2/3$ and $Q_d=-1/3$. Then we have
\vs\be
Q_{q_L} = \Bigl(t^3+ Y_{q_L}\Bigr) = \left(\ba{cc}
  +2/3&0\\0&-1/3\ea\right)~,
\label{qlcharge1}
\ee\vs\nin
which results in
\vs\be\boxed{
  Y_{q_L}=\frac{1}{6}~}~.
\label{qlhyp}
\ee\vs\nin
The hypercharge assignments for the right handed quarks are again
trivial and given by the quark electric charges. We have
\vs\be\boxed{
Y_{u_R} = +\frac{2}{3} ~,\qquad Y_{d_R} = -\frac{1}{3}~}~.
\label{rhqhyp}
\ee\vs\nin
With these hypercharge assignments we can now write the quark
couplings to the $Z$.  Using
(\ref{neutralfcoup2}) we obtain
\vs\bear
u_L: && \qquad -i\frac{g}{\cos\theta_W}\,\Bigl(\frac{1}{2}
-\sin^2\theta_W\frac{2}{3}\Bigr)~\nonumber\\
d_L: && \qquad -i\frac{g}{\cos\theta_W}\,\Bigl(-\frac{1}{2} +
\sin^2\theta_W\frac{1}{3}\Bigr)\nonumber\\
u_R: && \qquad
-i\frac{g}{\cos\theta_W}\,\Bigl(-\sin^2\theta_W\frac{2}{3}\Bigr)\nonumber\\
d_R: &&\qquad -i\frac{g}{\cos\theta_W}\,\Bigl(\sin^2\theta_W\frac{1}{3}\Bigr)~.
\label{qzcoup1}
\eear\vs\nin
Once again, we see that each type of quark has a different coupling to
the $Z$. All of these predictions have been tested with great
precision, confirming the SM even beyond leading order.

\nin
The charged gauged couplings of left handed quarks are trivial to obtain: they are
dictated by $SU(2)_L$ gauge symmetry and therefore there must be the
same as those of the left handed leptons in (\ref{chargedcoup2}). So we have
\vs\be
\cL_{q}^{\rm ch.} = \frac{g}{\sqrt{2}}\,\Bigl\{\bar u_L\gamma^\mu d_L \, W^+_\mu
+ \bar d_L\gamma^\mu u_L\,W^-_\mu\Bigr\}~.
\label{qchargedcoup}
\ee\vs\nin

\subsubsection{Fermion Masses} 

\nin
We have seen that SSB leads to masses for same of the gauge bosons,
preserving gauge invariance. We now direct our attention to fermion
masses. In principle these terms
\vs\be
\cL_{\rm fm} = m_\psi \bar \psi_L \psi_R +{\rm h.c.} ~,
\ee\vs\nin
are forbidden by $SU(2)L\times U(1)_Y$ gauge invariance since thy are
not invariant under
\vs\bear
\psi_L &\to & e^{i\alpha^a(x) t^a}\, e^{i\beta(x)
  Y_{\psi_L}}\,\psi_L\nonumber\\
~\label{fgaugetrans}
\psi_R &\to & e^{i\beta(x)
  Y_{\psi_R}}\,\psi_R ~.\nonumber
\eear\vs\nin
But the operator
\vs\be
\bar\psi_L\, \Phi \,\psi_R ~,
\ee\vs\nin
is clearly invariant under the $SU(2)_L$ gauge transformations, and it
would be $U(1)_Y$ invariant if\vs\be
-Y_{\psi_L} + Y_\Phi + Y_{\psi_R} =0~.
\label{hyperiszero}
\ee\vs\nin
Since $Y_\Phi=1/2$, this form of the oprator will work for the down
type quarks and charged leptons. For instance, since $Y_L=-1/2$ and
$Y_{e_R}=-1$, the operator
\vs\be
\cL_{me}=
\lambda_e \,\bar L \,\Phi \,e_R~ + {\rm h.c.},
\label{emass1}
\ee\vs\nin
is gauge invariant since the hypercharges satisfy (\ref{hyperiszero}).
In (\ref{emass1}) we defined the dimensionless coupling $\lambda_e$
which will result in a Yukawa coupling of electrons to the Higgs
boson. To see this, we write $\Phi(x)$ in the unitary gauge, so that
\vs\bear
\cL_{me} &=&\lambda_e\, (\bar\nu_{e_L}\quad \bar e_L)\, \left(\ba{c}0\\
  ~\\
  \frac{v+h(x)}{\sqrt{2}}\ea\right)\, e_R  + {\rm h.c.} \nonumber\\
&=&\lambda_e \frac{v}{\sqrt{2}}\,\bar e_L e_R +
\lambda_e\,\frac{1}{\sqrt{2}}\,h(x)\,\bar e_L e_R + {\rm h.c.}~,
\label{emass2}
\eear\vs\nin
where the first term is the electro mass term resulting in
\vs\be
m_e = \lambda_e\,\frac{v}{\sqrt{2}}~,
\label{emass3}
\ee\vs\nin
and the second term is the Yuawa interaction of the electron and the
Higgs boson $h(x)$. We can rewrite (\ref{emass2}) as
\vs\be
\cL_{\rm me} = 
m_e\,\bar e_L e_R +
\frac{m_e}{v}\,h(x)\,\bar e_L e_R + {\rm h.c.}~,
\label{emass4}
\ee\vs\nin
from which we can see that the electron couples to the Higgs boson
with a strength equal to its mass in units of the Higgs VEV
$v$. Similarly, for quarks we have that the operator
\vs\be
\cL_{md}=
\lambda_e \,\bar q_L \,\Phi \,d_R~ + {\rm h.c.},
\label{dmass1}
\ee\vs\nin
is gauge invariant since $Y_{q_L}=1/6$ and $Y_{d_R}=-1/3$ satisfy (\ref{hyperiszero}).
Them we obtain
\vs\be
\cL_{\rm md} = 
m_d\,\bar d_L d_R +
\frac{m_d}{v}\,h(x)\,\bar d_L d_R + {\rm h.c.}~, 
\label{dmass2}
\ee\vs\nin
and where the down quark mass was defined as
\vs\be
m_d = \lambda_d\,\frac{v}{\sqrt{2}}~.
\label{dmass3}
\ee\vs\nin
As we can see, it will be always the case that fermions couple to the
Higgs boson with the strength $m_\psi/v$. Thus, the heavier the
fermion, the stronger its coupling to the Higgs.

\nin
Finally, in order to have gage invariant operators with up type right handed
quarks we need to use the operator
\vs\be
\cL_{\rm mu} =  \lambda_u \, \bar q_L \tilde\Phi\,u_R + {\rm h.c.}~,
\label{umass1}
\ee\vs\nin
where we defined
\vs\be
\tilde\Phi(x) = i\sigma^2\,\Phi(x)^* = \left(\ba{c}
  \frac{v+h(x)}{\sqrt{2}}\\
  ~\\0\ea\right)~,
\label{tildephidef}
\ee\vs\nin
where in the last equality we are using the unitary gauge. It is
straightforward\footnote{Only need to use that $\sigma^2 \sigma^2=1$,
  and that $\sigma^2 (\sigma^a)^* \sigma^2 = -\sigma^a$.} to prove that $\tilde\Phi(x)$ is an $SU(2)_L$ doublet
with $Y_{\tilde\Phi}=-1/2$, which is what we need so as to make the
operator in (\ref{umass1}) invariant under $U(1)_Y$. Then we have
\vs\be
\cL_{\rm mu} = 
m_u\,\bar u_L u_R +
\frac{m_u}{v}\,h(x)\,\bar u_L u_R + {\rm h.c.}~, 
\label{umass2}
\ee\vs\nin
with
\vs\be
m_u = \lambda_u\,\frac{v}{\sqrt{2}}~.
\ee\vs\nin
The fermion Yukawa couplings are parameters of the SM. In fact, since
there are three families of quarks their Yuakawa couplings are in
general a non diagoinal three by three matrix. This fact has important
experimental consequences. On the other hand, we could imagina having
something similar if we introduce a right handed neutrino. This
however, might be beyond the SM, since this state does not have any
SM gauge quantum numbers.
\nin
Overall, the SM is determined by the paremeters $v, g, g' $ and
$\sin\theta_W$ in the electroweak gauge sector, plus all the Yukawa
couplings in the fermion sector leading to all the observed fermion
masses and mixings.

\section{Conclusions}
\label{sec:conclusions}

The standard model of particle physics is built by using a number of
tools in quantum field theory developed over decades. The experimental
guidance made possible to choose the correct tools. This has been  the case
with the SM gauge theory, $SU(3)_c\times SU(2)_L\times U(1)_Y$. The
couplings of these gauge bosons to matter have been tested with
increasing precision over the years. In the case of the electroweak
sector, it was also necessary to introduce a whole new sector, the
Higgs sector, in order to explain the origins of the massive gauge
bosons, the $W^\pm$ and $Z^0$. The Higgs mechanisms not only makes
these masses compatible with the $SU(2)_L\times U(1)_Y$ gauge
invariance, but also makes a number of precise predictions that have
been consistently confirmed by experiment. Then, the Higgs mechanism,
the spontaneous breaking of the electroweak symmetry leaving the
photon as the only massless electroweak gauge boson, gives a
consistent  picture of that completes the SM. Its last prediction, the
existence of a scalar boson, the Higgs, with the couplings predicted
by the Higgs mechanism, has also been confirmed experimentally.  

However, despite the extensive experimental tests confirming the
picture of the Higgs mechanism that we saw above, the SM leaves
several unanswered questions. Some of these questions are not meant to
be addressed by the SM. For instance, the origin of dark matter, or
that of the matter--anti-matter in the universe (which is zero in the
SM due to an accidental symmetry). Others, refer to the origin of the
dimensionless couplings in the SM, such as the values of the gauge
couplings or those of the Yukawa couplings of the fermions, the latter
leading to a rather large hierarchy of fermion masses. Even worse, the
SM does not have an origin for neutrino masses. But it is not clear
what is the energy scale where any of these question should be
answered. For instance, Yukawa couplings might be defined at very high
energies and then evolve to the electroweak scale by their
renormalization evolution. Or, in the case of dark matter or the
matter anti-matter asymmetry, the physics associated with it might be
out of our experimental reach.

On the other hand, the introduction of the Higgs sector in the SM
leading to the spontaneous breaking of the electroweak gauge symmetry,
requires the introduction of the {\em only} energy scale in the SM. In
fact, the mass term $m^2$ in (\ref{smpotential}) is the only dimensionfull quantity in
all of the SM lagrangian. It leads to the Higgs VEV  and to the masses
of all the SM particles\footnote{Perhaps with the exception of the
  neutrino masses.}. It is then a reasonable question to ask: what is
the origin of this mass scale ? Or more generally, what is the origin
of the Higgs sector in the SM. After all, the Higgs boson is the only
elementary scalar in the SM. As we saw in the first two sections, in
all other examples of spontaneous symmetry breaking in other physical
systems, the scalar field was a low energy effective degree of
freedom. Either it is a collective degree of freedom, as in condense
matter systems, or a composite one as in hadronic physics, where the
$\sigma$ hadron plays the role of the Higgs boson. Then, it is natural
to wonder if the Higgs boson is an elementary scalar or it is
composite.  A first attempt, called Technicolor, attempted to imitate
QCD by introducing a new set of fermions and a new interaction that
would lead to the formation of a ``techniquark'' condensate analogous
to the quark condensate in QCD. This would spontaneously break the
electroweak sector, resulting in masses for the $W $ and the
$Z$. Among the many problems of this kind of models, is that the Higgs
boson would be at the cutoff of the composite sector, just as the
$\sigma$ in QCD: it predicts a very massive ($m_h\simeq 1 TeV$) and
very broad Higgs boson, quite unlike what is observed. Another attempt
using the Nambu--Jona-Lasinio model to obtain a top quark condensate,
fares a bit better, but it still predicts a Higgs mass much too
large.

But there is a way to make a scalar like the Higgs boson
naturally light, and that is for it to be a (pseudo) Nambu-Goldstone
boson (pNGB). The models where this is the case are generically dubbed
composite Higgs models (CHM) and, once again, take a page from the low
energy strong interactions. There, although the $\sigma$  is heavy
with a mass at the cutoff of the theory, the pions are light since
they are the NGBs from the spontaneous breaking of the global chiral
symmetry $SU(2)_L\times SU(2)_R$. The reason why the pions are nos
massless is the {\em explicit} breaking of chiral symmetry, which in
the QCD lagrangian corresponds to non vanishing masses for the light
quarks. The same can be applied to the SM. Its gauge symmetry can be
embedded in a large global symmetry, which is then spontaneously
broken at a scale we generically call $f$, larger than the electroweak
scale $v$. The Higgs boson would be a massless NGB from this SSB of
the global symmetry. But since the symmetry is partially gauged by the
electroweak sector, i.e. by $SU(2)_L\times U(1)_Y$, this represents an
explicit breaking of the global symmetry. Also the Yukawa couplings of
fermions to the NGB Higgs further break the global symmetry
explicitly. As a result, the Higgs boson acquires a small mass
(compared to $f$ ) at one loop order.
Several models like this have been proposed in the literature. The
minimal symmetry breaking pattern that actually works is $SO(5)$,
spontaneously broken down to $SO(4)$, which implies the presence of 4
NGBs, with which the Higgs doublet is built. This model is called the
minimal composite Higgs model (MCHM), and it does not predict
additional particles at the weak scale $v$. However it does predict
additional states both bosonic and fermionic, at the scale $f$, the so
called resonances. These models provide examples of tests of the Higgs
sector of the SM which may illuminate both the nature of the SM scalar
sector as well as the origin of the electroweak scale. These tests
involve searching for the heavy resonances, as well as for the
predicted pattern of deviations in the Higgs couplings to the SM. 
The high luminosity LHC will be able to test the Higgs couplings with
great precision, although its energy reach for the resonance spectrum
will not be larger than the current one at the LHC. Thus, testing the
Higgs sector of the SM will involve both precision measurements and
predictions in order to obtain the most information on the possible
origin of the electroweak symmetry breaking scale.

\begin{ack}[Acknowledgments]%
 The author thanks the State of Sao Paulo Foundation for the Advancement of
 Research  (FAPESP), grant 2019/04837-9, for its support. 
\end{ack}


\bibliographystyle{Numbered-Style} 
\bibliography{reference}

\end{document}